\documentclass[10pt,conference,compsoc]{IEEEtran}

\ifCLASSOPTIONcompsoc
  \usepackage[nocompress]{cite}
\else
  \usepackage{cite}
\fi
\usepackage{wrapfig}
\usepackage{alltt}
\usepackage{amsmath}
\usepackage{amssymb}
\usepackage{graphicx}
\usepackage{color}
\usepackage{hyperref}
\usepackage{cleveref}
\usepackage{hhline}

\newtheorem{corollary}{Corollary}
\newtheorem{definition}{Definition}

\newtheorem{lemma}{Lemma}

\newtheorem{remark}{Remark}

\newtheorem{theorem}{Theorem}

\Crefname{theorem}{Thm.}{Thm.}
\Crefname{section}{\S}{\S\S}
\Crefname{definition}{Def.}{Def.}
\Crefname{figure}{Fig.}{Fig.}
\Crefname{appendix}{App.}{App.}

\newcommand{\remspace}{\vspace{-0.2cm}}

\renewcommand{\vec}[1]{\overline{#1}}

\newcommand{\proj}[2]{{{#1}|_{#2}}}

\newcommand{\Nat}{{\mathbb{N}}}

\newcommand{\true}{{\textit{true}}}

\newcommand{\vocabulary}{{\ensuremath{\Sigma}}}

\newcommand{\struct}{s}
\newcommand{\Dom}{{\mathcal{D}}}
\newcommand{\Int}{{\mathcal{I}}}

\newcommand{\propsem}{P}

\newcommand{\states}{{S}}
\newcommand{\initsem}{{S_0}}
\newcommand{\trsem}{{R}}

\newcommand{\init}{\iota}
\newcommand{\tr}{{\tau}}

\newcommand{\initm}{{\widetilde{\varphi_0}}}
\newcommand{\trm}{{\tilde{\tau}}}

\newcommand{\state}{s}

\newcommand{\sort}[1]{{\textit{#1}}}
\newcommand{\sthread}{\sort{thread}}

\newcommand{\snumber}{\sort{number}}

\newcommand{\relation}[1]{{\textit{#1}}}
\newcommand{\rscheduled}{{\relation{scheduled}}}
\newcommand{\ridle}{{\relation{idle}}}
\newcommand{\rwait}{{\relation{wait}}}
\newcommand{\rcrit}{{\relation{critical}}}

\newcommand{\G}{{\square}}
\newcommand{\F}{{\lozenge}}

\newcommand{\sharon}[1]{{\textcolor{blue}{SH: {\em #1}}}}
\newcommand{\oded}[1]{{\textcolor{cyan}{OP: {\em #1}}}}
\newcommand{\TODO}[1]{{\textcolor{red}{TODO: {\em #1}}}}
\newcommand{\commentout}[1]{}
\newcommand{\OMIT}[1]{}
\newcommand{\DONE}[1]{{\textcolor{green}{DONE: {\em #1}}}}

\renewcommand{\sharon}[1]{}
\renewcommand{\oded}[1]{}
\renewcommand{\TODO}[1]{}
\renewcommand{\DONE}[1]{}

\newcommand{\remove}[1]{\xspace}

\newcommand{\allnotes}[1]{}
\newcommand{\para}[1]{\vspace{3pt}{\em #1.}}

\newcommand{\tsystem}{{T}}
\newcommand{\insystem}{{\tsystem_S}}
\newcommand{\tabsystem}{{\tsystem_A}}
\newcommand{\productsystem}{{\tsystem_P}}
\newcommand{\witnesssystem}{{\tsystem_W}}

\newcommand{\tabvocab}{{\vocabulary_A}}
\newcommand{\productvocab}{{\vocabulary_P}}
\newcommand{\witnessvocab}{{\vocabulary_W}}

\newcommand{\productinit}{{\init_P}}
\newcommand{\witnessinit}{{\init_W}}

\newcommand{\tabtr}{{\tr_A}}
\newcommand{\producttr}{{\tr_P}}
\newcommand{\witnesstr}{{\tr_W}}

\newcommand{\witconsts}{C}

\newcommand{\relg}[1]{{r_{\G {#1}}}}
\newcommand{\assgn}{\sigma}
\newcommand{\tstate}{s}
\newcommand{\sat}[1]{{\operatorname{FO}\left[{#1}\right]}}
\newcommand{\A}{{A}}
\newcommand{\B}{{B}}

\newcommand{\footprintstate}[1]{{f({#1})}}

\newcommand{\footprinti}[1]{{f(\state_0,\ldots,\state_{#1})}}

\newcommand{\redproves}[2]{#1 \vdash #2}

\newcommand{\safety}{\sigma}

\begin{document}

%
% paper title
% Titles are generally capitalized except for words such as a, an, and, as,
% at, but, by, for, in, nor, of, on, or, the, to and up, which are usually
% not capitalized unless they are the first or last word of the title.
% Linebreaks \\ can be used within to get better formatting as desired.
% Do not put math or special symbols in the title.

%\title{Proving temporal properties with temporal invariants}
%\title{Temporal Prophecy for Infinite-State Temporal Verification}
\title{Temporal Prophecy for Proving Temporal Properties of Infinite-State Systems}

%% Authors:
%% Oded Padon, Jochen Hoenicke, Kenneth L. McMillan, Andreas Podelski, Mooly Sagiv and Sharon Shoham

% author names and affiliations
% use a multiple column layout for up to three different
% affiliations
%% \author{\IEEEauthorblockN{Michael Shell}
%% \IEEEauthorblockA{School of Electrical and\\Computer Engineering\\
%% Georgia Institute of Technology\\
%% Atlanta, Georgia 30332--0250\\
%% Email: http://www.michaelshell.org/contact.html}
%% \and
%% \IEEEauthorblockN{Homer Simpson}
%% \IEEEauthorblockA{Twentieth Century Fox\\
%% Springfield, USA\\
%% Email: homer@thesimpsons.com}
%% \and
%% \IEEEauthorblockN{James Kirk\\ and Montgomery Scott}
%% \IEEEauthorblockA{Starfleet Academy\\
%% San Francisco, California 96678-2391\\
%% Telephone: (800) 555--1212\\
%% Fax: (888) 555--1212}}

% conference papers do not typically use \thanks and this command
% is locked out in conference mode. If really needed, such as for
% the acknowledgment of grants, issue a \IEEEoverridecommandlockouts
% after \documentclass

% for over three affiliations, or if they all won't fit within the width
% of the page (and note that there is less available width in this regard for
% compsoc conferences compared to traditional conferences), use this
% alternative format:
%
\author{\IEEEauthorblockN{Oded Padon\IEEEauthorrefmark{1},
Jochen Hoenicke\IEEEauthorrefmark{2},
Kenneth L. McMillan\IEEEauthorrefmark{3},
Andreas Podelski\IEEEauthorrefmark{2},
Mooly Sagiv\IEEEauthorrefmark{1} and
Sharon Shoham\IEEEauthorrefmark{1}}
\IEEEauthorblockA{\IEEEauthorrefmark{1}Tel Aviv University, Israel~~~~
\IEEEauthorrefmark{2}University of Freiburg, Germany~~~~
\IEEEauthorrefmark{3}Microsoft Research, USA}}

%% \author{\IEEEauthorblockN{Michael Shell\IEEEauthorrefmark{1},
%% Homer Simpson,
%% James Kirk, 
%% Montgomery Scott and
%% Eldon Tyrell}
%% \IEEEauthorblockA{\IEEEauthorrefmark{1}School of Electrical and Computer Engineering\\
%% Georgia Institute of Technology,
%% Atlanta, Georgia 30332--0250\\ Email: see http://www.michaelshell.org/contact.html}}

%% \author{\IEEEauthorblockN{Michael Shell\IEEEauthorrefmark{1},
%% Homer Simpson\IEEEauthorrefmark{2},
%% James Kirk\IEEEauthorrefmark{3}, 
%% Montgomery Scott\IEEEauthorrefmark{3} and
%% Eldon Tyrell\IEEEauthorrefmark{4}}
%% \IEEEauthorblockA{\IEEEauthorrefmark{1}School of Electrical and Computer Engineering\\
%% Georgia Institute of Technology,
%% Atlanta, Georgia 30332--0250\\ Email: see http://www.michaelshell.org/contact.html}
%% \IEEEauthorblockA{\IEEEauthorrefmark{2}Twentieth Century Fox, Springfield, USA\\
%% Email: homer@thesimpsons.com}
%% \IEEEauthorblockA{\IEEEauthorrefmark{3}Starfleet Academy, San Francisco, California 96678-2391\\
%% Telephone: (800) 555--1212, Fax: (888) 555--1212}
%% \IEEEauthorblockA{\IEEEauthorrefmark{4}Tyrell Inc., 123 Replicant Street, Los Angeles, California 90210--4321}}

% make the title area
\maketitle

\begin{abstract}

% motivation
Various verification techniques for temporal properties transform temporal verification to safety verification.
For infinite-state systems, these transformations are inherently imprecise.
That is, for some instances, the temporal property holds, but the resulting safety property does not.
This paper introduces a mechanism for tackling this imprecision.
This mechanism, which we call \emph{temporal prophecy}, is inspired by prophecy variables.
Temporal prophecy refines an infinite-state system using first-order linear temporal logic formulas,
via a suitable tableau construction.
For a specific liveness-to-safety transformation based on first-order logic, %, where temporal prophecy is defined via first-order linear temporal logic,
we show that using temporal prophecy strictly increases the precision.
Furthermore, temporal prophecy leads to robustness of the proof method, which is manifested by a cut elimination theorem.
We integrate our approach into the Ivy deductive verification system, and show that it can handle challenging temporal verification examples.

\end{abstract}

% For peer review papers, you can put extra information on the cover
% page as needed:
% \ifCLASSOPTIONpeerreview
% \begin{center} \bfseries EDICS Category: 3-BBND \end{center}
% \fi
%
% For peerreview papers, this IEEEtran command inserts a page break and
% creates the second title. It will be ignored for other modes.
\IEEEpeerreviewmaketitle

\section{Introduction}
\label{sec:intro}
\remspace

%\subsection{Introduction}

%In this paper, we introduce a proof technique that uses temporal
%prophecy to improve the precision of abstractions that reduce liveness
%proofs to safety proofs.
% \sharon{the abstractions don't reduce. Maybe "abstractions used for reducing...}

There are various techniques in the literature that transform the
problem of verifying liveness of a system to the problem of verifying
safety of a different system. These transformations compose the system
with a device that has the known property that some safety condition
$\safety$ implies liveness. The classical example of this is proving
termination of a while loop with a ranking function.  In this case,
the device evaluates a chosen function $r$ on loop entry, where 
the range of $r$ is a well-founded set. The safety property $\safety$
is that $r$ decreases at every iteration, which implies that the loop
must terminate.

A related transformation, due to Armin Biere~\cite{livenessassafety},
applies to finite-state (possibly parameterized) systems. The safety
property $\safety$ is, in effect, that no state occurs twice, from
which we can infer termination. In the infinite-state case, this can
be generalized using a function~$f$ that projects the program state
onto a finite set.  We can think of this as a ranking that tracks the
set of unseen values of $f$ and is ordered by set inclusion. However,
the property that no value of $f$ occurs twice is simpler to verify,
since the composed device can non-deterministically guess the
recurring value. In general, the effectiveness of a liveness-to-safety
transformation depends strongly on the difficulty of the resulting
safety proof problem.

Other methods can be seen as instances of this general approach. For
example, the Terminator tool~\cite{cook-terminator} might be seen as
combining the ranking and the finite projection approaches.  Another
approach by Fang \emph{et al.} applies a collection of ad-hoc devices
with known safety-to-liveness properties to prove liveness of
parameterized protocols~\cite{DBLP:conf/forte/FangMPZ06}.  Of greatest
interest here, a recent paper by Padon \emph{et al.} uses a
dynamically chosen finite projection that depends on a finite prefix
of the system's execution~\cite{popl18}. The approach
of~\cite{DBLP:conf/cav/ManevichDR16} also has some similar
characteristics.

In the case of infinite-state systems, these transformations from
liveness verification to safety verification are not precise
reductions. That is, while safety implies liveness, a counterexample
to the safety property $\safety$ does not in general imply a
counterexample to liveness. For example, in the projection method, a
terminating infinite-state system may have runs whose length exceeds
the finite range of any chosen projection $f$, forcing some value to
repeat.

In this paper, we show that the precision of a liveness-to-safety
transformation can be usefully increased by the addition of
\emph{prophecy variables}. These variables are expressed as
first-order LTL formulas.  For example, suppose we augment the state
of the system with a variable $\relg{p}$ that tracks the truth value
of the proposition $\G p$, which is true when $p$ holds in all future
states.  We can soundly add two constraints to the transition
system. To the transition relation, we add $\relg{p} \leftrightarrow
(p \wedge \relg{p}')$, where $\relg{p}'$ denotes the value of the
prophecy variable in the post-state. We also add the fairness
constraint that $\relg{p} \vee \neg p$ holds infinitely often. These
constraints are typical of tableau constructions that convert a
temporal formula to a symbolic automaton.  As we show in this paper,
the additional information they provide refines the trace set of the
transformed system, potentially eliminating false counterexamples.

In particular, we will show how to integrate tableau-based prophecy
with the liveness-to-safety transformation of~\cite{popl18} that uses
a history-based finite projection, referred to as \emph{dynamic abstraction}.
We show that the precision of this transformation is consequently increased.
The result is that we can prove properties that otherwise would not be directly provable using the technique.

% \subsubsection{Temporal prophecy in general}

% The example was described for a particular temporal prophecy formula.
% The paper generalizes this idea to arbitrary prophecy formulas in FO-LTL, using an infinite-state symbolic tableau construction. As demonstrated by the example, temporal prophecy has the potential to eliminate spurious abstract lassos.
% As a result, the refined reduction can prove strictly more properties and systems, as we will observe in several case studies.

% Although temporal prophecy does not make the liveness to safety
% reduction complete, we show that the set of systems and properties
% that can be proven via the reduction with temporal prophecy has an
% elegant closure property (\Cref{thm:closure}).  In particular, it
% admits a form of \emph{cut elimination}: if the reduction can prove
% that some system satisfies both $\varphi \to \psi$ and $\neg \varphi
% \to \psi$, then it can also prove the system satisfies $\psi$.

% %\paragraph{Implementation}
% We have implemented the liveness to safety reduction with temporal prophecy in the Ivy deductive verification system~\cite{ivy}.
% Our approach allows a user friendly language in which the user can state the invariant for the safety property resulting from the reduction. We evaluate our approach on examples that  require temporal prophecy (and cannot be proven without it):
% the ticket protocol with task queues, the alternating bit protocol, and the TLB shootdown protocol.

%\subsubsection{Main contributions}
%\subsubsection*{Main contributions}

This paper makes the following contributions:
\begin{enumerate}
%\item Define temporal prophecy using FO-LTL formulas and temporal witnesses using
\item Introduce the notion of temporal prophecy, including prophecy formulas and prophecy witnesses, \emph{via} a first-order LTL tableau construction.
\TODO{is it really only via the tableau? don't the witnesses use a different scheme?}\oded{I think we can say both are via the tableau}
\item Show that temporal prophecy increases the proof power (i.e., precision) of the safety-to-liveness transformation based on dynamic abstraction, and further show that the properties provable with temporal prophecy are closed under \emph{first-order reasoning}, with cut elimination as a special case. % of this closure.
\item Integrate the liveness-to-safety transformation based on dynamic abstraction and temporal prophecy into the Ivy deductive verification system, deriving the prophecy formulas from an inductive invariant provided by the user (for proving the safety property).
\item Demonstrate the effectiveness of the approach on some challenging examples that cannot be handled by the transformation without temporal prophecy.
\item Demonstrate that prophecy witnesses can eliminate quantifier alternations in the verification conditions generated for the safety problem obtained after the transformation, facilitating decidable reasoning.
\end{enumerate}

\section{Illustrative Example}
\label{sec:intro:ticket}
\remspace

\begin{figure}
%\vspace{-1cm}
\begin{center}
\vspace{0.5cm}
\includegraphics[scale=0.5]{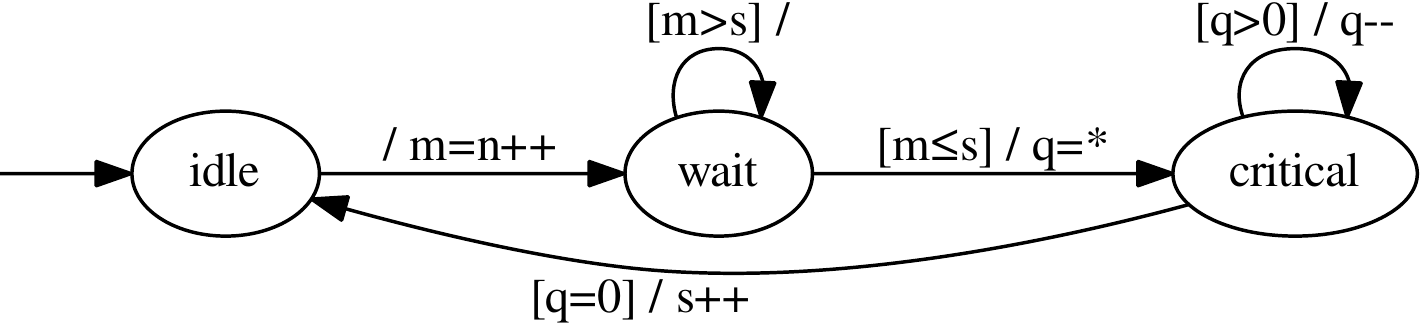}
\vspace{-2.0cm}
\begin{scriptsize}
\begin{alltt}
global nat s, n
local nat m, q
\end{alltt}
\end{scriptsize}
\vspace{0.5cm}
\end{center}
\caption{\label{fig:ticket}%
\TODO{improve caption}
The ticket mutual exclusion protocol.
%Threads are either idle, waiting to enter the critical section, or in the critical section.
Edges are labeled by condition~/~action.
}
%\vspace{-0.5cm}
\end{figure}
%\end{wrapfigure}

We illustrate our approach using the ticket protocol for ensuring mutual exclusion with non starvation
among multiple threads, depicted in \Cref{fig:ticket}. The ticket protocol may be run by any number of threads, and also allows dynamic spawning of threads.
The protocol is an idealized version of spinlocks used in the Linux kernel~\cite{spinlocks}. In the protocol, each thread can be in one of three states: idle, waiting to enter the critical section, or in the critical section. The right to enter the critical section is determined by a ticket number.
A global variable $n$, records the next available ticket,
and a global variable $s$, records the ticket currently being served.
Each thread has a local variable $m$ that records the ticket it holds.
A thread only enters the critical section when $m \leq s$.
Once a thread enters the critical section, it handles tasks that accumulated in its task queue, and stays in the critical section until its queue is empty (tasks are only added to the queue when the thread is outside the critical section).
In \Cref{fig:ticket}, this is modeled by the task counter $q$, a thread-local variable which is non-deterministically set when a thread enters the critical section (to account for the unbounded, but finite, number of tasks), and is then decremented in each step. When $q = 0$ the thread leaves the critical section, and increments $s$ to allow other threads to be served.

%We note that the ticket protocol may be run by any number of
%threads, and also allows dynamic spawning of threads.

%This is a ticket-based mutual exclusion protocol in which each process takes the next available ticket number $n$, incrementing $n$, then waits for all processes with lesser tickets to exit (see~\Cref{fig:ticket}).

The protocol is designed to satisfy the following first-order temporal property:
\[\left(\forall x. \G \F \rscheduled(x) \right) \to \forall y. \G \left(\rwait(y) \to \F \rcrit(y) \right)\]
That is, if every process is scheduled infinitely often, then every waiting process eventually enters its critical section.
(Note that we encode fairness assumptions as part of the temporal property.)

\para{Insufficiency of liveness-to-safety transformations}
While the temporal property is clearly satisfied by the ticket
protocol, proving it is challenging for liveness-to-safety transformations.
First, due to the unbounded values obtained by the ticket number and
the task counter, and also due to dynamic spawning of threads, this
example does not belong to the class of parameterized
systems~\cite{DBLP:conf/cav/PnueliS00}, where a simple lasso argument is sound
(and complete) for proving liveness. Second, while using a finite abstraction can recover soundness,
no fixed finite abstraction is precise enough
to show the absence of a lasso-shaped counterexample in this example.
The reason is that a thread can go to
the waiting state ($\rwait$) with any number of threads waiting
``ahead of it in line''.
%
%The reason is that there is no fixed $N$ such that the eventuality must hold after at most $N$ steps (where
%in each step, all fairness constraints are satisfied at least once): a thread $t$ can go to
%the waiting state ($\rwait$) with any number $k$ of threads waiting
%``ahead of it in line'', i.e., with smaller ticket numbers. This will
%require $k$ steps to ensure granting $t$'s request to enter the
%critical section.  As a result, any finite abstraction will still include cycles stemming from traces where the eventuality takes longer than the number of states the abstraction can distinguish.
%
%The reason is that a finite abstraction can show an eventuality only if there is a fixed
%$N$ such that the eventuality must hold after at most $N$ steps (where
%in each step, all fairness constraints are satisfied at least once).
%If no such $N$ exists, any finite abstraction is bound to include cycles stemming from traces where the eventuality takes longer than the number of states the abstraction can distinguish.
%Unfortunately, in the ticket protocol, there is no such $N$, since a thread $t$ can go to
%the waiting state ($\rwait$) with any number $k$ of threads waiting
%``ahead of it in line'', i.e., with smaller ticket numbers. This will
%require $k$ steps to ensure granting $t$'s request to enter the
%critical section.

For cases where no finite abstraction is sufficiently precise to prove liveness, we may
instead apply the liveness-to-safety transformation of~\cite{popl18}.  This
transformation relaxes the requirement of proving absence of lassos over a fixed finite abstraction,
and instead requires one to prove absence of lassos over a \emph{dynamic} finite abstraction that is only determined after some prefix of the trace (allowing for better precision). Soundness is maintained since the abstraction is still finite.
Technically, the technique requires to prove that no \emph{abstract lasso} exists, where
%a more restricted class of counterexamples, referred to as \emph{abstract lasso}.
%
%uses a \emph{dynamic abstraction} that depends on a (prefix) of the trace.
%Namely, the reduction requires one to prove absence of a more restricted class of counterexamples, referred to as \emph{abstract lasso}.
%
%This transformation restricts the form of a counterexample to an
%\emph{abstract lasso}.
an abstract lasso is a finite execution prefix
that (i)~visits a \emph{freeze point}, at which a finite projection (abstraction) of
the state space is fixed, (ii) the freeze point is followed by two
states that are equal in the projection. We refer to these as the
\emph{repeating states}, and (iii)~all fairness constraints are
visited both before the freeze point and between the repeating states.

%Because the projection (or abstraction) is determined at the freeze point and not fixed in advance,
%we refer to the abstraction as \emph{dynamic}.
Unlike fixed finite abstractions, dynamic abstractions allow us to prove
that an eventuality holds if there is a finite upper bound on the number of
steps required \emph{at the time the eventuality is asserted} (the freeze point). The
bound need not be fixed \emph{a priori}.
%% Unfortunate, for the ticket protocol the dynamic abstraction is still insufficiently precise, since it
%% fixes the projection only at the time when $t$ makes its request. It does not require an \emph{a priori} fixed bound on the number of steps required for the request to be granted,
%% %by fixing the abstraction only at the time when $t$ makes its request, and can therefore overcome the issue described above.
%% but it still requires such a bound to exist at the time of the request.
%% %However, a bound must exist at the time of the request.
Unfortunately, due to the non-determinism introduced by the task
counter $q$, each of the $k$ threads ahead of $t$ in line could
require an unbounded number of steps to leave the critical section,
and this number is not yet determined when $t$ makes its request.  As
a result, there is an abstract lasso which freezes the abstraction
when $t$ makes its request, after which some other thread $t_0$ enters
the critical section and loops, decrementing its task counter $q$.
Since the value of the task counter of $t_0$ is not captured in the
abstraction, the loop does not change the abstract state.
%% with a value of $q$ that is not
%% captured by the abstraction.  The repeating states will be two states
%% such that all threads have been scheduled between them, but the only
%% thread that changed its state is $t_0$, which decreased its value of
%% $q$. However, these two states have the same projection.
This spurious abstract lasso prevents this liveness-to-safety transformation from proving the property.

\para{Temporal prophecy to the rescue}
The key to fixing this problem is to predict the future to the extent
that a bound on the steps required for progress is determined at the
freeze point. Surprisingly, this is accomplished by the use of one temporal prophecy variable corresponding to the truth value of the following formula:
\[\exists x. \F \G \rcrit(x).\]
If this formula is initially true, there is some thread $t_0$ that eventually enters the critical section and stays there.
At this point, we can prove it eventually exits (a contradiction) because the number of steps needed for this is bounded by the current task counter of $t_0$.
Operationally, the freeze point is delayed until $\G \rcrit(x)$ holds at which point $t_0$'s task counter is captured in the finite projection, ruling out an abstract lasso.
On the other hand if the prophecy variable is initially false, then all threads are infinitely often out of the critical section. With this fairness constraint,
thread $t$ requires only a finite number of steps to be served, determined by the number of threads with lesser tickets. Operationally, the extra fairness constraint extends the lasso loop until the abstract state must change, ruling out an abstract lasso.

\commentout{
Though the abstraction cannot handle the problem as given, by
splitting the initial states on the value of a future-time formula, we
obtain two cases that it \emph{can} handle. In both cases, the
abstraction is refined by additional fairness conditions.  In one case,
this results in postponing the freeze point, thus capturing more
information in the abstraction, while in the other it rules out
looping behaviors.
\sharon{this paragraph shouts "case splitting" -- a point we do not want to make. I think we should rephrase it:}
}

Though the liveness-to-safety transformation via dynamic abstraction
and abstract lasso detection cannot handle the problem as given,
introducing suitable temporal prophecy eliminates the spurious
abstract lassos.  Some spurious lassos are eliminated by postponing
the freeze point, thus refining the finite abstraction, and others are
eliminated by additional fairness constraints on the lasso loop.
This example is explained in greater detail in \Cref{sec:ticket}.

\section{Preliminaries}
\label{sec:prelim}
\remspace

In this section, we present the first-order formalism for specifying infinite-state systems and their properties, as well as a tableau construction for first-order LTL formulas.

\subsection{Transition Systems in First-Order Logic}
\label{sec:fofts}
\remspace

A first-order logic transition system is a triple $(\vocabulary, \init, \tr)$, where
$\vocabulary$ is a first-order vocabulary that contains only relation symbols and constant symbols (functions can be encoded by relations),
$\init$ is a closed formula over $\vocabulary$ defining the set of initial states, and $\tr$ is
a closed formula over $\vocabulary \uplus \vocabulary'$, where $\vocabulary' = \{\ell' \mid \ell \in \vocabulary\}$, defining the transition relation. The constants in $\vocabulary$ represent the program variables.

A state of the transition system is a first-order structure, $\state = (\Dom, \Int)$, over $\vocabulary$, where $\Dom$ denotes the (possibly infinite) domain of the structure and $\Int$ denotes the interpretation function.
The set of initial states is the set of all states $\state$ such that $\state \models \init$, and the set of transitions is the set of all pairs of states $(\state,\state')$ with the same domain such that $(\struct,\struct') \models \tr$.
In the latter, $(\struct,\struct')$ denotes a structure over the vocabulary $\Sigma \uplus \Sigma'$ with the same domain as $\struct$ and $\struct'$ in which the symbols in $\Sigma$ are interpreted as in $\struct$, and the symbols in $\Sigma'$ are interpreted as in $\struct'$.
%In the latter, given $\struct = (\Dom,\Int)$ and $\struct' = (\Dom,\Int')$, we use $(\struct,\struct')$ as a shorthand for the structure $(\Dom,\Int \uplus \Int'')$, where
%$\Int'' = \lambda r' \in \Sigma'. \Int'(r)$. Namely, the structure defined by $(\struct,\struct')$ is a structure over the vocabulary $\Sigma \uplus \Sigma'$ with the same domain as $\struct$ and $\struct'$, and where the symbols in $\Sigma$ are interpreted as in $\struct$, and the symbols in $\Sigma'$ are interpreted as in $\struct'$.
%Namely, the structure defined by $(\struct,\struct')$ is a structure over the vocabulary $\Sigma \uplus \Sigma'$ with the same domain as $\struct$ and $\struct'$, and where the symbols in $\Sigma$ are interpreted as in $\struct$, and the symbols in $\Sigma'$ are interpreted as in $\struct'$.

\commentout{
For a state $\state = (\Dom,\Int)$ over $\vocabulary$, and for $\Dom' \subseteq \Dom$ that contains the interpretation of all the constants, we denote by $\proj{\state}{\Dom'}$ the state obtained by projecting $\state$ to $\Dom'$, i.e., $\proj{\state}{\Dom'}=(\Dom', \Int')$, where for every constant symbol $c \in \vocabulary$, $\Int'(c) = \Int(c)$ and for every relation symbol $r \in \vocabulary$ of arity $k$, $\Int'(r) = \Int(r) \cap (\Dom')^k$.
}\TODO{rethink this for the final version}
For a state $\state = (\Dom,\Int)$ over $\vocabulary$, and for $D \subseteq \Dom$,
we denote by $\proj{\state}{D}$ the partial structure by projecting $\state$ to $D$, i.e.,
$\proj{\state}{D}=(D, \proj{\Int}{D})$,
where $\proj{\Int}{D}$ interprets only constants $c \in \vocabulary$ for which $\Int(c) \in D$ (making it a partial interpretation),
and for every relation symbol $r \in \vocabulary$ of arity $k$, $\proj{\Int}{D}(r) = \Int(r) \cap D^k$.
For a vocabulary $\vocabulary' \subseteq \vocabulary$, we denote by $\proj{\state}{\vocabulary'}$ the state over $\vocabulary'$ obtained by restricting the interpretation function to the symbols in $\vocabulary'$, i.e., $\proj{\state}{\vocabulary'} = (\Dom, \Int')$, where for every symbol $\ell \in \vocabulary'$, $\Int'(\ell) = \Int(\ell)$.

A (finite or infinite) \emph{trace} of $(\vocabulary, \init, \tr)$ is a sequence of states $\pi = \state_0,\state_1,\ldots$ where $\state_0 \models \init$ and $(\state_i, \state_{i+1}) \models \tr$ for every $0 \leq i < |\pi|$.
Every state along the trace has its own interpretation of the constant and relation symbols,
but they all share the same domain. % $\Dom$.
%The reachable states of $(\Sigma, \init, \tr)$ consist of all the states that reside on a trace of $(\vocabulary, \init, \tr)$.
%reachable in $(\states, \initsem, \trsem)$ for some $\Dom$.

%We now provide a formalism for specifying transition systems in first-order logic.
We note that first-order transition systems are  Turing-complete. Furthermore, tools such as Ivy~\cite{ivy} provide modeling languages that are closer to imperative programming languages and compile to a first-order transition system. This makes it easier for a user to provide a first-order specification of the transition system they wish to verify.
%We use the standard syntax and semantics of first-order logic.

\para{Safety} %\TODO{define safety property as one that has finite counterexamples}
Given a vocabulary $\vocabulary$, a safety property $\propsem$ is a set of sequences of states over $\vocabulary$, such that for every sequence of states $\pi \not \in \propsem$, there exists a finite prefix $\pi'$ of $\pi$, such that $\pi'$ and all of its extensions are not in $\propsem$. % \not\in \propsem$ and every extension of $\pi'$ is also not in $\propsem$.
A transition system over $\vocabulary$ satisfies $\propsem$ if all of its traces are in $\propsem$.

\subsection{First-Order Linear Temporal Logic (FO-LTL)}
\remspace

To specify temporal properties of first-order transition systems we use First-Order Linear Temporal Logic (FO-LTL), which combines LTL with first-order logic~\cite{abadi89-power}.
For simplicity, we consider only the ``globally'' ($\G$) temporal operator. The tableau construction extends to other operators as well, and so does our approach. % can be extended to handle other operators as well.
%Other temporal operators, such as ``eventually'' ($\F$), can be expressed using $\G$. \TODO{Our approach can be extended to handle other operators as well.}

\para{Syntax}
Given a first-order vocabulary $\Sigma$, FO-LTL formulas are
%\oded{removing ``the closed formulas'', since they are not all closed}
%obtained by the following grammar:
defined by:
\commentout{
\[
%f :: = \varphi \mid \neg f \mid f_1 \lor f_2 \mid \exists x. f \mid \forall x. f \mid \X f \mid f_1 \U f_2
 f :: = r(t_1,\ldots,t_n) \mid t_1 = t_2 \mid \neg f \mid f_1 \lor f_2 \mid \exists x. f %\mid \forall x. f
\mid \G f
\qquad  t :: =  c \mid x
\]
}
%\begin{small}
\begin{align*}
%f :: = \varphi \mid \neg f \mid f_1 \lor f_2 \mid \exists x. f \mid \forall x. f \mid \X f \mid f_1 \U f_2
 f & :: = r(t_1,\ldots,t_n) \mid t_1 = t_2 \mid \neg f \mid f_1 \lor f_2 \mid \exists x. f %\mid \forall x. f
\mid \G f \\
t & :: =  c \mid x
\end{align*}
%\end{small}
where $r$ is an $n$-ary relation symbol in $\Sigma$, $c$ is a constant symbol in $\Sigma$, $x$ is a variable,
each $t_i$ is a term over $\Sigma$  %(defined as in first-order logic)
and $\G$ denotes the ``globally'' temporal operator.
We also use the standard shorthand for the ``eventually'' temporal operator: $\F f = \neg \G \neg f$, and the usual shorthands for logical operators (e.g., $\forall x. f  = \neg \exists x. \neg f$).
% \longversion{
%\begin{align*}
%& \F f = \textit{true} \ \U f
%& \G f = \neg \F \neg f
%\end{align*}
%}

\para{Semantics}
FO-LTL formulas over $\vocabulary$ are interpreted over infinite sequences of states (first-order structures) over $\vocabulary$.
%of a first-order transition system $(\Sigma, \init, \tr)$  (with the same vocabulary).
%\footnote{We interpret FO-LTL formulas over transition systems without fairness since fairness can be specified as part of the FO-LTL formula: all fair traces of $(\Sigma, \init, \tr, \fs)$ satisfy $f \in$ FO-LTL iff all  traces of $(\Sigma, \init, \tr)$ satisfy $(\bigwedge_{\f \in \fs} \forall \bar{x}.\,\G\F\f(\bar{x})) \to f$.}.
%
Atomic formulas are interpreted over states, % (which are first-order structures),
the temporal operators are interpreted as in traditional LTL, and first-order quantifiers are interpreted over the shared domain $\Dom$ of all states in the trace.
%
%Recall that each trace of such a system is a sequence of first-order structures that share the same domain $\Dom$, but may interpret the constant and relation symbols differently.
%%The first-order quantifiers are therefore interpreted over elements of $\Dom$, the temporal operators are interpreted as in traditional LTL, and the non-temporal first-order formulas are interpreted over individual states (structures).
%\oded{changed this a bit} The first-order quantifiers are therefore interpreted over elements of $\Dom$, and the temporal operators are interpreted as in traditional LTL.
%%In particular, note that different states may interpret the constant and relation symbols differently, and hence may interpret first-order formulas differently.
%
Formally, the semantics is defined w.r.t. an infinite sequence of states $\pi=\state_0,\state_1,\ldots$ and an assignment $\assgn$ that maps variables to $\Dom$ --- the shared domain of all states in $\pi$. We define $\pi^i = \state_i,\state_{i+1},\ldots$ to be the suffix of %the sequence
$\pi$ starting at index $i$.
The semantics is defined as follows.
\begin{align*}
\pi, \assgn \models r(t_1,\ldots,t_n) & \Leftrightarrow \state_0, \assgn \models r(t_1,\ldots,t_n) \\
\pi, \assgn \models t_1 = t_2 & \Leftrightarrow \state_0, \assgn \models t_1 = t_2 \\
\pi, \assgn \models \neg \psi & \Leftrightarrow \pi, \assgn \not \models \psi \\
\pi, \assgn \models \psi_1 \lor \psi_2 & \Leftrightarrow \pi, \assgn \models \psi_1 \text{ or } \pi,\assgn \models \psi_2 \\
\pi, \assgn \models \exists x. \psi & \Leftrightarrow \text{exists $d \in \Dom$ s.t. } \pi, \assgn[x \mapsto d] \models \psi \\
\pi, \assgn \models \G \psi & \Leftrightarrow \text{forall $i \geq 0$, } \pi^i, \assgn \models \psi
\end{align*}
When the formula has no free variables, we omit $\assgn$.
%For a formal definition of FO-LTL semantics see~\cite{abadi89-power}.
%
A first-order transition system $(\Sigma, \init, \tr)$ satisfies a closed FO-LTL formula $\varphi$ over $\Sigma$ if all of its traces satisfy $\varphi$. % \longversion{(for every $\Dom$).}

%Note that the fairness constraints $\fs$ of the transition system can always be specified as part of the FO-LTL formula:
%$(\Sigma, \init, \tr, \fs)$ satisfies $f$ if and only if $(\Sigma, \init, \tr)$ satisfies $\left(\bigwedge_{\f \in \fs} \forall \bar{x}.\,\G\F\f(\bar{x})\right) \to f$.

\subsection{Tableau for FO-LTL}
\label{sec:tableau}
\remspace

As part of our liveness-to-safety transformation, we use a standard tableau construction for FO-LTL formulas that results in a first-order transition system with \emph{fairness constraints}.
Unlike the classical construction, we define the tableau for a set of formulas, not necessarily a single temporal formula.
%(and its subformulas).
%We include the construction and some of its properties for completeness of the presentation.

%\subsection{Tableau Construction}

%\TODO{maybe "universal tableau" -- check if "universal" already has meaning}

%augment a transition system with

For an FO-LTL formula $\varphi$, we denote by $sub(\varphi)$ the set of subformulas of $\varphi$, defined in the usual way.
In the sequel, we consider a finite set $\A$ of FO-LTL formulas that is closed under subformulas, i.e. for every $\varphi \in \A$, $sub(\varphi) \subseteq \A$.
Note that $\A$ may contain formulas with free variables.

\oded{I made more things into labeled definitions, to address the review: ``In the paper I did not find the definition of tableau (what is exactly meant by this term). From the context in section 3.3 I got the impression that tableau is for a set of formulas A is the transition system $T_A$ but I would like to see that stated explicitly.'' Could you please check and see that it makes sense?}

%% Given a finite set $\A$ as above over a first-order vocabulary $\vocabulary$, %transition system $(\vocabulary,\init,\tr)$
%% we extend $\vocabulary$ to $\tabvocab$ by introducing a relation symbol $\relg{\varphi}$ of arity $k$ for every formula $\G \varphi \in \A$ with $k$ free variables.
\begin{definition}[Tableau vocabulary]
Given a finite set $\A$ as above over a first-order vocabulary $\vocabulary$,
the \emph{tableau vocabulary} for $\A$, denoted $\tabvocab$, is obtained from $\vocabulary$
by adding a fresh relation symbol $\relg{\varphi}$ of arity $k$ for every formula $\G \varphi \in \A$ with $k$ free variables.
\end{definition}
Recall that $\G$ is the only primitive temporal operator we consider (a similar construction can be done for other operators).
The symbols added in $\tabvocab$ will be used to ``label''  states by temporal subformulas that are satisfied by all outgoing fair traces. To translate temporal formulas over $\vocabulary$ to first-order formulas over $\tabvocab$ we use the following definition.
%% We then define a first-order transition system $\tabsystem = (\tabvocab, \true, \tabtr)$ in which the states are ``labeled'' (via the new symbols in $\tabvocab \setminus \vocabulary$) by temporal subformulas that are satisfied by all of their outgoing fair traces.

%
%Given a first-order vocabulary $\vocabulary$ %transition system $(\vocabulary,\init,\tr)$
%and a finite set $\A$ as above over $\vocabulary$, we define a first-order transition system $\tabsystem = (\tabvocab, \true, \tabtr)$.
%Intuitively speaking, the states in $\tabsystem$ are ``labeled'' (via the new symbols in $\tabvocab \setminus \vocabulary$) by temporal subformulas that are satisfied by all of their outgoing fair traces.
%
%Formally, $\tabvocab$ is defined to extend $\vocabulary$ with a relation $\relg{\varphi}$ of arity $k$ for every formula $\G \varphi \in \A$ with $k$ free variables. Recall that $\G$ is the only primitive temporal operator we consider (a similar construction can be done for other operators).

%% In order to define the transition relation and the notion of a fair trace, we need the following definition.

\begin{definition}
For a FO-LTL formula $\varphi \in \A$ (over $\vocabulary$), its first-order representation, denoted $\sat{\varphi}$, is a first-order formula over $\tabvocab$, defined inductively, as follows.
\begin{align*}
\sat{\varphi} &= \varphi \text{\ \ \  if $\varphi = r(t_1,\ldots,t_n)$ or $\varphi= t_1 = t_2$} \\
\sat{\G \psi (\vec{x})} &= \relg{\psi (\vec{x})}(\vec{x}) \\
\sat{\neg \psi} &= \neg \sat{\psi} \\
\sat{\psi_1 \lor \psi_2} &= \sat{\psi_1} \lor \sat{\psi_2}\\
\sat{\exists x. \psi} &= \exists x. \sat{\psi}
\end{align*}
%$\sat(\G \psi (\vec{x})) = \relg{\psi}$.
\end{definition}
Note that $\sat{\varphi}$ has the same free variables as $\varphi$.
We can now define the tableau for $\A$ as a transition system.

\begin{definition}[Tableau transition system]
The \emph{tableau transition system} for $\A$ is the first-order transition system
$\tabsystem = (\tabvocab, \true, \tabtr)$, where $\tabtr$ (defined over $\tabvocab \uplus  \tabvocab'$) is defined as follows:
\[\tabtr = \bigwedge_{\G \varphi \in \A} \forall \vec{x}. \; (\relg{\varphi}(\vec{x}) \leftrightarrow (\sat{\varphi(\vec{x})} \land \relg{\varphi}'(\vec{x}))).\]
\end{definition}
Note that the original symbols in $\vocabulary$ (and $\vocabulary'$) are not constrained by $\tabtr$, and may change arbitrarily with each transition. However, the $\relg{\varphi}$ relations are updated in accordance with the property that $\pi,\assgn \models \G p$ iff $\state_0,\assgn \models p$ and $\pi^1,\assgn \models \G p$ (where $\state_0$ is the first state of $\pi$ and $p$ is a first-order formula over $\vocabulary$).

%Intuitively speaking, for a closed FO-LTL formula $\G \varphi \in \A$, whenever a state satisfies the relation $\relg{\varphi}$, all of its fair outgoing traces satisfy $\G \varphi$, and whenever the state is not labeled by $\relg{\varphi}$, all of its fair outgoing traces satisfy $\neg \G \varphi$.

\begin{definition}[Fairness] \label{def:fairness}
%\sharon{I didn't call it "fair trace" since it may not be a trace (we don't want to commit to a certain transition system, since first we talk about $\tabsystem$ but in the theorem we talk about the product.}
A sequence of states $\pi = \tstate_0, \tstate_1,\ldots $ over $\tabvocab$ is $\A$-\emph{fair} if for every temporal formula $\G \varphi(\vec{x}) \in \A$ and for every assignment $\assgn$,
there are infinitely many $i$'s for which
$\tstate_i, \assgn \models \sat{ \G \varphi(\vec{x}) \lor \neg \varphi(\vec{x})}$. % for infinitely many $i$'s.
%\relg{\varphi}(\vec{x}) \lor \neg \varphi(\vec{x})$  holds for infinitely many $i$'s.
\end{definition}
\oded{added this, please check:}Note that $\G \varphi(\vec{x}) \lor \neg \varphi(\vec{x})$, used above, is equivalent to $\F \neg\varphi(\vec{x}) \to \neg \varphi(\vec{x})$. So the definition of fairness ensures an eventuality cannot be postponed forever.
In the sequel, the set $\A$ is always clear from the context (e.g., from the vocabulary), hence we omit it and simply say that $\pi$ is fair.
%When $\A$ is clear from the context, we omit it and simply say that $\pi$ is fair.

The next claims summarize the properties of the tableau; \Cref{lem:tableau-only-satisfying traces} states that the FO-LTL formulas over $\vocabulary$ that hold in the outgoing traces of a tableau state correspond to the first-order formulas over $\tabvocab$ that hold in the state; \Cref{lem:tableau-all-satisfying traces} states that every sequence of states over $\vocabulary$ has a representative trace in the tableau; finally, \Cref{thm:tab} states that a transition system satisfies a FO-LTL formula iff its product with the tableau of the negated formula has no fair traces.
\begin{lemma} \label{lem:tableau-only-satisfying traces}
In a fair trace $\pi = \tstate_0, \tstate_1,\ldots $ of $\tabsystem$ (over $\tabvocab$), for every FO-LTL formula %$\G \varphi(\vec{x}) \in \A$,
$\psi(\vec{x}) \in \A$,
%$I_i(\relg{\varphi}) = \{ (e_1,\ldots,e_k) \mid \pi^i,(e_1,\ldots,e_k) \models \G \varphi \}$.
%Alternatively,
for every assignment $\assgn$ and for every index $i \in \Nat$, we have that
$\tstate_i,\assgn \models \sat{\psi(\vec{x})}$ %\relg{\varphi}(\vec{x})$
iff $\pi^i,\assgn \models \psi(\vec{x})$. %\G \varphi(\vec{x})$.
%where $\pi^i = \tstate_i,\tstate_{i+1},\ldots$ denotes the suffix of $\pi$ starting at $\tstate_i$.
\end{lemma}

\begin{lemma} \label{lem:tableau-all-satisfying traces}
Every infinite sequence of states $\hat{\state}_0, \hat{\state}_1,\ldots $ over $\vocabulary$ can be extended to a fair trace $\pi = \tstate_0, \tstate_1,\ldots $ of $\tabsystem$ (over $\tabvocab$)
s.t. for every $i \in \Nat$, $\proj{\tstate_i}{\vocabulary} = \hat{\state}_i$.
%the domain of $\tstate_i$ and $\hat{\state}_i$ is the same, and they agree on their interpretations of the symbols in $\vocabulary$.
%$\proj{\tstate_i}{\vocabulary} = \hat{state}_i$.
 %\sharon{abuse of notation (projection): here projecting to vocabulary, not domain}
% over $\tabvocab$.
\end{lemma}
%\begin{IEEEproof}
%\TODO{Construct the fair trace.}
%\end{IEEEproof}

\begin{definition}[Product system] \label{def:product}
Given a transition system %$\tsystem
$\insystem = (\vocabulary, \init,\tr)$, a closed FO-LTL formula $\varphi$ over $\vocabulary$,
a finite set $\A$ of FO-LTL formulas over $\vocabulary$ closed under subformulas such that $\neg \varphi \in \A$,
we define the \emph{product system} of $\insystem$ and $\neg \varphi$ over $\A$ as the first-order transition system
$\productsystem = (\productvocab, \productinit, \producttr)$
given by $\productvocab = \tabvocab$, $\productinit = \init\wedge \sat{\neg \varphi}$ and $\producttr = \tr \wedge \tabtr$,
where $\tabsystem = (\tabvocab,\true,\tabtr)$ is the tableau for $\A$.

%Given a transition system %$\tsystem
%$\insystem = (\vocabulary, \init,\tr)$ and a finite set of FO-LTL formulas closed under subformulas $\A$, %\sharon{do we also need that $\A$ is closed under subformula?}
%the \emph{product system} of $\insystem$ and $\A$ is the first-order transition system
%$\productsystem = (\productvocab, \productinit, \producttr)$, where
%$\productvocab = \tabvocab$, $\productinit = \init\wedge \sat{\neg \varphi}$ and $\producttr = \tr \wedge \tabtr$, where $\tabsystem = (\tabvocab,\true,\tabtr)$ is the tableau for $\A$.
\end{definition}

\begin{theorem} \label{thm:tab}
Let $\productsystem$ be the product system of $\insystem$ and $\neg \varphi$ over $\A$ as defined in \Cref{def:product}.
%Given a transition system %$\tsystem
%$\insystem$ % = (\vocabulary, \init,\tr)$
%and a closed FO-LTL formula $\varphi$, let $\A$ be a finite set of FO-LTL formulas closed under subformulas such that $\neg \varphi \in \A$,
%and let $\productsystem$ be the product system of $\insystem$ and $\A$.
Then $\insystem \models \varphi$ iff $\productsystem$ has no fair traces.
%
%and let $\tabsystem = (\tabvocab, \true, \tabtr)$ be the tableau for $\A$.
%Then $\insystem \models \varphi$ iff the transition system $(\tabvocab, \init\wedge \sat{\neg \varphi}, \tr \wedge \tabtr)$ has no fair traces.
%%$\product{\tsystem}{\tableauts{\A}} = (\tabvocab, \init\wedge \sat{\neg \varphi}, \tr \wedge \tabtr)$ has no fair traces.
\end{theorem}

%We will sometimes refer to $(\tabvocab, \init\wedge \sat{\neg \varphi}, \tr \wedge \tabtr)$ as a \emph{product system}.
Intuitively, the product system augments the states of $\insystem$ with temporal formulas from $\A$,
splitting each state into many (often infinitely many) states according to the future behavior of its outgoing traces.
Note that \Cref{thm:tab} holds already when $\A = sub(\neg \varphi)$. However, as we will see, taking a larger set $\A$ is useful for proving fair termination via the liveness-to-safety transformation.

\section{Liveness-to-Safety with Temporal Prophecy}
\label{sec:reduction}
\remspace

%\sharon{should we keep the new order here? first describe the acyclicity condition (safety property) and only later explain the augmentation with prophecy?}

In this section we present our liveness proof approach using temporal
prophecy and a liveness-to-safety transformation. As in earlier
approaches, our transformation (i) uses a tableau construction to construct
a product transition system equipped with fairness constraints such
that the latter has no fair traces iff the temporal property holds of
the original system, and (ii) defines a safety property over the
product transition system such that safety implies that no fair traces
exist (note that the opposite direction does not hold).

The gist of our liveness-to-safety transformation is that we augment the construction of the product transition system with two forms of prophecy detailed in \Cref{sec:prophecy}.
We then use the definition of the safety property from~\cite{popl18}. In the sequel, we first present the safety property and then present the augmentation with temporal prophecy,
whose goal is to ``refine'' the product system such that it will be safe.

\subsection{Safety Property: Absence of Abstract Lassos}
\label{sec:safety}
\remspace

%The essence of the reduction of the problem of showing absence of a fair execution to safety verification~\cite{popl18} is defining a
%safety property that ensures that a (fair) transition system has no fair executions.
%%a (product) transition system $\witnesssystem$ over $\tabvocab$ (e.g., as constructed in \Cref{thm:tab}) has no fair executions (in the context of \Cref{thm:tab}, this means that the original transition system satisfies its temporal property).

Given a transition system $\witnesssystem = (\witnessvocab,\witnessinit, \witnesstr)$ with $\witnessvocab \supseteq \tabvocab$ (e.g., the product system from \Cref{def:product}), we define a notion of an abstract lasso, whose absence is a safety property that implies that $\witnesssystem$ has no $\A$-fair traces. % and hence that $\insystem \models \varphi$.
This section recapitulates material from~\cite{popl18}.

The definition of an abstract lasso is based on a dynamic abstraction that is fixed at some point along the trace, henceforth called the \emph{freeze point}.
The abstraction function is defined by projecting a state (a first-order structure) into a finite subset of its domain.
This finite subset is defined by the union of the \emph{footprints} of all states encountered until the freeze point,
where the footprint of a state includes the interpretation it gives all constants from $\witnessvocab$. % (which include constants for prophecy witnesses).
Intuitively, the footprint includes all elements ``exposed'' in the state, including those ``touched'' by outgoing transitions.

%We start by defining an abstraction function, based on the \emph{footprint} of a finite trace.

%%\para{Footprint}
%Our dynamic abstraction is defined by projecting a state (a first-order structure) into a finite set of elements from its domain.
%This finite subset is defined by the union of the \emph{footprints} of all states encountered until the point in which the abstraction is fixed, henceforth called the \emph{freeze point}.
%Intuitively, the footprint of a state will include the interpretation it gives all constants from $\witnessvocab$ (which include constants for prophecy witnesses).
%%For transitions, the footprint of every transition includes the interpretation of constants in $\vocabulary$ in the post-state.
%%This captures the values of all program variables after the transition.
%%(note that only constants from the original vocabulary represent program variables that are mutable, so these are the only ones included in the footprint of transitions).

\begin{definition}[Footprint]
For a state $\state = (\Dom,\Int)$ over $\witnessvocab$, we define %s.t. $\state \models \init_W$, define
the footprint of $\state$ as $\footprintstate{\state} = \{\Int(c) \mid c \in \witnessvocab\}$.
%Given $\state, \state'$ s.t. $\state,\state' \models \tr_W$, define
%$\footprinttr{\state}{\state'} = \{\Int_{\state'}(c) \mid c \in \vocabulary\}$.
For a sequence of states $\pi = \state_0,\state_1,\ldots$ over $\witnessvocab$, and an index $i < |\pi|$, we define the footprint of $\state_0,\ldots,\state_i$ as
$\footprinti{i} = \bigcup_{j=0}^{i} \footprintstate{\state_j}$.
\end{definition}
Importantly, the footprint of a finite trace is always finite. As a result, an abstraction function that maps each state to the result of projecting it to the footprint of the trace until the freeze point has a finite range.

\begin{definition}[Fair Segment]
Let $\pi = \state_0,\state_1,\ldots$ be a sequence of states over $\witnessvocab$.
For $0 \leq i \leq j < |\pi|$, we say the segment $[i,j]$ is fair if
for every formula $\G \psi(\vec{x}) \in \A$, and for every assignment $\assgn$ where every variable is assigned to an element of $\footprinti{i}$,
there exists $i \leq k \leq j$ s.t. $\state_k,\assgn \models \sat{ (\G \psi(\vec{x})) \lor \neg \psi(\vec{x})}$.
\end{definition}

\commentout{
Formally, this is captured by the following initial and update formulas for $d$:

\begin{align*}
\initm &= waiting \land \neg frozen \land \neg saved \land
\exists \vec{x}, \vec{y}^1,\ldots,\vec{y}^m. initmatrix(\vec{x}) \land
\left( \bigwedge_{i} \forall \vec{z}. \psi_i(\vec{z}) \to \psi_i(\vec{y}_i) \right) \land \\
&\forall z. d(z) \leftrightarrow \left(
\bigvee_{c \in \vocabulary} z=c \lor
\bigvee_i z=x_i \lor
\bigvee_{j,k} z=y_j^k
\right) \land
\bigwedge_i \left( \forall \vec{z}. w_i(\vec{z}) \leftrightarrow \bigwedge_j d(z_j) \right)
\end{align*}

\begin{align*}
\trm &= \exists \vec{x}, \vec{y}^1,\ldots,\vec{y}^m. trmatrix(\vec{x}) \land
\left( \bigwedge_{i} \forall \vec{z}. \psi'_i(\vec{z}) \to \psi'_i(\vec{y}_i) \right) \land \\
&\forall z. d'(z) \leftrightarrow \left( d(z) \lor
\bigvee_{c \in \vocabulary} z=c' \lor
\bigvee_i z=x_i \lor
\bigvee_{j,k} z=y_j^k
\right) \land
% update w
%\bigwedge_i \left( \forall \vec{z}. w_i(\vec{z}) \leftrightarrow \bigwedge_j d(z_j) \right)
\end{align*}
}

%\begin{definition}
%An $\A$-abstract cycle is $s_f,s_1,s_2$ s.t.:
%1) For all $\G \varphi$ in $\A$, for any tuple in footprint(s_{init})
%2)
%\end{definition}

%\para{Abstract Lassos} % and the Reduction}

\begin{definition}[Abstract Lasso]
A finite trace $\state_0,\ldots,\state_n$  of $\witnesssystem$  is an \emph{abstract lasso}
if there are $0 \leq i \leq j < k \leq n$ s.t.
the segments $[0,i]$ and $[j,k]$ are fair, and
$\proj{\state_j}{\footprinti{i}} = \proj{\state_k}{\footprinti{i}}$.
\end{definition}
Intuitively, in the above definition, $i$ is the \emph{freeze point}, where the abstraction is fixed. % to map each state $\state$ later on in the trace to its projection to $\footprinti{i}$.
The states $\state_j$ and $\state_k$ are the ``repeating states'' -- states that are indistinguishable by the abstraction that projects them to the footprint $\footprinti{i}$.
The segment between $j$ and $k$, respectively, the segment between $0$ and $i$, meet all the fairness constraints restricted to elements in $\footprinti{j}$, respectively, in $\footprintstate{\state_0}$.
Fairness of the segment $[0,i]$ is needed to prevent the freeze point from being chosen too early, thus creating spurious abstract lassos.
Note that the absence of abstract lassos is a safety property. %, which completes our reduction.

%\para{Reduction}

\begin{lemma}
If $\witnesssystem$ has no abstract lassos then it also has no fair traces.
%If $\witnesssystem$ has a fair trace, then it also has an abstract lasso.
\end{lemma}
\begin{IEEEproof}
Assume to the contrary that $\witnesssystem$ has a fair trace $\pi = \state_0,\state_1,\ldots$.
Let $i$ be the first index such that $[0,i]$ is fair (such an index must exist since the set
$\footprintstate{\state_0}$, which determines the relevant fairness constraints is finite).
Since $\footprinti{i}$ is also finite, there must exist an infinite subsequence $\pi'$ of $\pi^i$ such that for every $\state,\state'$ in this subsequence $\proj{s}{\footprinti{i}} = \proj{s'}{\footprinti{i}}$. Let $j \geq i$ be the index in $\pi$ of the first state in $\pi'$. $\footprinti{j}$ is also finite, hence there exists $k' > j$ such that the segment $[j,k']$ of $\pi$ is fair. Take $k$ to be the index in $\pi$ of the first state of $\pi^{k'}$ that is also in $\pi'$. Since $\pi'$ is infinite, such a $k$ must exist. Since $k \geq k'$, the segment $[j,k]$ is also fair. This defines an abstract lasso $s_0,\ldots,s_i,\ldots,s_j,\ldots,s_k$, in contradiction.
\end{IEEEproof}

%Note that the absence of abstract lassos is a safety property. %, which completes our reduction.

\subsection{Augmenting the Transition System with Temporal Prophecy}
\label{sec:prophecy}
\remspace
%The reduction of the problem of checking that $\witnesssystem$ has no fair traces to the problem of checking absence of abstract lassos in inevitably incomplete.
%This makes the overall reduction of the problem of checking temporal properties of $\insystem$ to the problem of safety verification incomplete.
%We now show how to ``improve'' the completeness of the reduction with the help of temporal prophecy that allows to construct a more refined product system $\witnesssystem$.
%\TODO{better phrasing?}

In this section we explain how our liveness-to-safety transformation
constructs $\witnesssystem = (\witnessvocab,\witnessinit,
\witnesstr)$, to which we apply the safety property of
\Cref{sec:safety}.  Our construction exploits both \emph{temporal
  prophecy formulas} and \emph{prophecy witnesses}, explained below.
\commentout{
First, the tableau construction explained in \Cref{sec:tableau} is used to augment the original transition system with a set $\A$ of FO-LTL formulas that contains not only the negation of the verified property (and its subformulas), but may also contain additional temporal prophecy formulas.

Second, a set $\B \subseteq \A$ is used to augment the transition system with fresh constants that serve as \emph{prophecy witnesses} for existential properties.
Next, we explain this in more detail. %the construction of the augmented transition system in two steps.
}
For the rest of this section we fix a first-order transition system $\insystem = (\vocabulary,\init,\tr)$ and a closed FO-LTL formula $\varphi$ over $\vocabulary$ that we wish to verify in $\insystem$.

\para{Temporal Prophecy Formulas}
First, given a set $\A$ of (not necessarily closed) FO-LTL formulas closed under subformula that contains $\neg \varphi$,
we construct the product system $\productsystem = (\productvocab, \productinit, \producttr)$ defined in \Cref{def:product}.
By \Cref{thm:tab}, $\insystem \models \varphi$ iff $\productsystem$ has no fair traces.
Note that classical tableau constructions are defined with $\A = sub(\neg\varphi)$, and we allow $\A$ to include more formulas.
These formulas act as ``temporal prophecy variables'' in the sense that they split the states of $\insystem$, according to the future behavior of outgoing traces.

While the liveness-to-safety transformation is already sound with $\A = sub(\neg \varphi)$, one of the chief observations of this work is that temporal prophecy formulas improve its precision.
These additional formulas in $\A$ split the states of $\insystem$ into more states in $\productsystem$,
and they cause some non-determinism of the future trace to be ``pulled backwards''
(the outgoing traces contain less non-determinism).
For example, if $\relg{\varphi}$ holds for some elements in the current state,
then $\varphi$ must continue to hold for these elements in the future of the trace.
Similarly, for elements where $\relg{\varphi}$ does not hold, there will be some time in the future of the trace where $\varphi$ would not hold for them.

This is exploited by the liveness-to-safety transformation in three ways, eliminating spurious abstract lassos.
First, having more temporal formulas in $\A$ refines the definition of a fair segment, and postpones the freeze point,
thus making the abstraction defined by the footprint up to the freeze point more precise.
For example, if $\relg{\varphi}$ does not hold for a ground formula $\varphi$ in the initial state, then the freeze point would be postponed until after $\varphi$ does not hold for the first time.
Second, it strengthens the requirement on the looping segment $s_j\ldots s_k$, in a similar way.
Third, the additional relations in $\productvocab (=\tabvocab)$ are part of the state as considered by the transformation,
and a difference in these relations (projected to the footprint up to the freeze point) is a valid difference.
These three ways all played a role in the examples considered in our evaluation.

\commentout{
\sharon{ what do we want to do with this? I moved it from the prelim, since any important intuition should be part of the reduction section}
The tableau construction reduces reasoning about FO-LTL to reasoning
about fair traces of the augmented transition system. To gain a more
intuitive understanding of the meaning of the new relations, consider
the fact that every state of the original system, corresponds to many
states in the augmented system that interpret the symbols in
$\vocabulary$ in the same way, but provide different interpretations
to the symbols in $\vocabof{\A}$. These interpretations differ by
allowing different sets of fair traces that leave the
state. Therefore, a state in the augmented system contains some
information not only about the current state, but also about the
future of the trace. For example, a state where $\relg{\varphi}$ holds
for some elements, already contains the information that $\varphi$
continues to hold for these elements in the entire future of the
trace. Similarly, for elements where $\relg{\varphi}$ does not hold,
there will be some time in the future of the trace where $\varphi$
would not hold for them.
}

\TODO{explain it also adds more relations on which we can discover a difference}

\TODO{maybe explain how this ``pulls the non-determinism backwards''}

\TODO{explain analogy to SMT}

%Given $(\init\wedge \sat{\neg \varphi}, \tr \wedge \tr_{tableau(\\A)})$.

%\subsection{Step 2: Abstract Lasso Detection with Prophecy Witnesses}
%%\subsection{Prophecy Witnesses}
%Our starting point for the second step of the reduction is the first-order transition system $\product{\tsystem}{\tableauts{\A}} = (\vocabof{\A}, \init\wedge \sat{\neg \varphi}, \tr \wedge \tableautr{\A})$, for which we need to show that there are no fair traces.

\para{Prophecy Witnesses}
The notion of an abstract lasso, used to define the safety property, considers a finite abstraction according to the footprint, which depends on the constants of the vocabulary.
%Namely, the interpretation of the constants defines a finite subset of the domain that is the basis for the abstraction.
To increase the precision of the abstraction,
%and also to support the cut elimination theorem (see \Cref{thm:closure}), \oded{is this correct?}
we augment the vocabulary with fresh constants that serve as \emph{prophecy witnesses} for existential properties.
%we use a set $\B \subseteq \A$ to augment the transition system with fresh constants that serve as \emph{prophecy witnesses} for existential properties.

To illustrate the idea, consider the formula $\psi(x) = \F \G p(x)$ where $x$ is a free variable.
If $\psi$ holds for some element, it is useful to include in the vocabulary a constant that serves as a witness for $\psi(x)$, and whose interpretation will be taken into account by the abstraction.
If $\psi$ holds for some $x$, the interpretation of the constant will be taken from such an $x$. Otherwise, this constant will be allowed to take any value.
%in case it holds for some $x$.
%In case that there is some $x$ s.t. $\psi(x)$ holds, it is useful to include in the vocabulary a witness to such an element.
%This constant will be allowed to take any value in case no such $x$ exists.
%\sharon{Remove: This refines the abstraction, and also lets the user refer to such an element in the inductive invariant. }
\TODO{mention Hilbert's choice operator}

%\begin{remark}
Temporal prophecy witnesses not only refine the abstraction,
they can also be used in the inductive invariant.
In particular, as demonstrated in the TLB Shootdown example (see \Cref{sec:eval}), in some cases this allows to avoid quantifier alternation cycles in the verification conditions, leading to decidability of VC checking.
%\end{remark}

Formally, %the second step of the reduction to safety is parameterized
%by the set $\A$ used in the tableau construction, as well as
given a set $\B \subseteq \A$, we construct $\witnesssystem = (\witnessvocab,\witnessinit, \witnesstr)$ as follows.
We extend $\productvocab$ to $\witnessvocab$ by adding fresh constant symbols $c_1,\ldots,c_n$ for every formula $\psi(x_1,\ldots,x_n) \in \B$.
We denote by $\witconsts$ the set of new constants, i.e., $\witnessvocab = \productvocab \cup \witconsts$.
The transition relation formula is extended to keep the new constants unchanged, i.e. $\witnesstr = \producttr \wedge \bigwedge_{c \in \witconsts} c = c'$, and
we define $\witnessinit$ by
%conjoining $\productinit$ with the following formula:
%
\begin{align*}
\witnessinit = \productinit \wedge \sat{
\left(\exists x_1,\ldots,x_n. \psi(x_1,\ldots,x_n)\right) \to
\psi(c_1,\ldots,c_n)
}
\end{align*}
Namely, $c_1,\ldots,c_n$ are required to serve as witnesses for $\psi(x_1,\ldots,x_n)$ in case it holds in the initial state for some elements, and otherwise they may get any interpretation at the initial state, after which their interpretation remains unchanged.
Adding these fresh constants and their defining formulas to the initial state is a conservative extension,
in the sense that every fair trace of $\productsystem$ can be extended to a fair trace of $\witnesssystem$ (fairness of traces over $\witnessvocab \supseteq \tabvocab$ is defined as in \Cref{def:fairness}), and every fair trace of $\witnesssystem$ can be projected to a fair trace of $\productsystem$.
As such we have the following:
%as formalized in the following lemma.

\begin{lemma}
Let $\productsystem = (\productvocab, \productinit, \producttr)$ and $\witnesssystem = (\witnessvocab,\witnessinit, \witnesstr)$ be defined as above.
Then $\productsystem$ has no fair traces iff $\witnesssystem$ has no fair traces.
%
%Let $\product{\tsystem}{\tableauts{\A}} = (\vocabof{\A}, \init\wedge \sat{\neg \varphi}, \tr \wedge \tableautr{\A})$ be defined as above.
%If $\B = \left\{ \psi_1,\ldots,\psi_k\right\}$ is a finite set of FO-LTL formulas over $\vocabulary$, and
%$c_{i,j} \not \in \vocabulary$ are fresh constant symbols,
%then $\product{\tsystem}{\tableauts{\A}}$ has no fair traces iff
%$(\vocabof{\A}\cup\{c_{i,j}\}, \init\wedge \sat{\neg \varphi \land
%\bigwedge_{i=1}^{k} \left(\exists x_1,\ldots,x_{n_i}. \psi(x_1,\ldots,x_{n_i})\right) \to \psi(c_{i,1},\ldots,c_{i,n_i})
%}, \tr \wedge \tableautr{\A})$ has no fair traces.
\end{lemma}

%$T_{S,\varphi,A,B} = (\vocabulary_{S,\varphi,A,B}, \init_{S,\varphi,A,B}, \tr_{S,\varphi,A,B})$
%
%$T_W(S,\varphi,A,B) = (\vocabulary_W, \init_W, \tr_W)$
%
%$T_W =(\vocabulary_W, \init_W, \tr_W)$

%\begin{lemma}
%Given closed $\varphi$ in FO-LTL, $\A$ be a finite set that includes all subformulas of $\varphi$,
%$\B = \left\{ \psi_1,\ldots,\psi_k\right\}$ be a finite set of FO-LTL formulas, and
%$c_{i,j}$ be fresh constant symbols,
%then $(\init,\tr) \models \varphi$ iff
%$(\init\wedge \sat{\neg \varphi \land
%\bigwedge_{i=1}^{k} \left(\exists x_1,\ldots,x_{n_i}. \psi(x_1,\ldots,x_{n_i})\right) \to \psi(c_{i,1},\ldots,c_{i,n_i})
%}, \tr \wedge \tr_{tableau(\A)})$ has no fair traces.
%\end{lemma}

%Given a first-order transition system $\tsystem = (\vocabulary,\init,\tr)$, a closed FO-LTL formula $\varphi$ %$\A$ be a finite set that includes all subformulas of $\varphi$,
%a finite set $\B = \left\{ \psi_1,\ldots,\psi_k\right\}$ of FO-LTL formulas, and fresh constant symbols
%$c_{i,j} \not \in \vocabulary$,
%we have that $(\vocabulary,\init,\tr) \models \varphi$ iff
%$(\init\wedge \sat{\neg \varphi \land
%\bigwedge_{i=1}^{k} \left(\exists x_1,\ldots,x_{n_i}. \psi(x_1,\ldots,x_{n_i})\right) \to \psi(c_{i,1},\ldots,c_{i,n_i})
%}, \tr \wedge \tr_{tableau(\A)})$ has no fair traces.

The overall soundness of the liveness-to-safety transformation is given by the following theorem.
\begin{theorem}[Soundness]
\label{thm:soundness}
Given a first-order transition system $\insystem$ and a closed FO-LTL formula $\varphi$ both over $\vocabulary$,
and given a set of temporal prophecy formulas $\A$ over $\vocabulary$ that contains $\neg \varphi$ and is closed under subformula,
and a set of temporal prophecy witness formulas $\B \subseteq \A$, if $\witnesssystem$ as defined above
does not contain an abstract lasso, then $\insystem \models \varphi$.
\end{theorem}

% \TODO{Remark: another example where this is helpful: we know $\forall x. \F \varphi(x)$, and need to wait until $\exists x. \varphi(x)$.}

\subsection{The Ticket Example}
\label{sec:ticket}

In this section we show in greater detail how prophecy increases
the power of the liveness-to-safety transformation.
%consider in greater detail how prophecy can increase
%the power of liveness-to-safety constructions, from a practical and a
%theoretical point of view.
%\subsection{Nested Ticket Example}
%\label{sec:ticket}
%\TODO{prove that sub-formulas of the property are insufficient}
As an illustration we return to the ticket example of \Cref{fig:ticket}.
As explained in \Cref{sec:intro:ticket}, in this example the liveness-to-safety transformation without temporal prophecy fails
%(see \Cref{app:ticket-fail} for a detailed proof),
(similarly to \cite[\S 5.2]{popl18}),
but it succeeds when adding suitable temporal prophecy.
%On the other hand, when adding the temporal prophecy formula $\exists x. \F \G \rcrit(x)$ to the tableau construction, no abstract lasso exists in the augmented transition system, and reduction succeeds to prove the property.

%and explain
%how the reduction succeeds with temporal prophecy. \Cref{} provides a detailed proof for why the reduction fails
%without it.

To model the ticket example as a first-order transition system, we use
a vocabulary with two sorts: $\sthread$ and $\snumber$.  The first
represents threads, and the second represents ticket values and
counter values.  The vocabulary also includes a static binary relation
symbol $\leq : \snumber,\snumber$, with suitable first-order axioms to
make it a total order.  (for more details about modeling systems in
first-order logic see e.g. \cite{ivy}.)  The state of the system is
modeled by unary relations for the program counter:
$\ridle,\rwait,\rcrit$, constant symbols of sort $\snumber$ for the
global variables $n,s$, and functions from $\sthread$ to $\snumber$
for the local variables $m,c$. The vocabulary also includes a unary
relation $\rscheduled$, which holds the last scheduled thread.

Next we show that when adding the temporal prophecy formula $\exists x. \F \G \rcrit(x)$ to the tableau construction, no abstract lasso exists in the augmented transition system, hence the liveness-to-safety transformation succeeds to prove the property.
Formally, in this case, $\A$ includes the following two formulas and their subformulas:
\begin{footnotesize}
\begin{align*}
& \neg \left( \left(\exists x. \neg \G \neg\G\neg \rscheduled(x) \right) \lor \neg \exists x. \neg \G \left(\neg\rwait(x) \lor \neg\G\neg \rcrit(x) \right) \right) \\
& \exists x. \neg\G\neg \G \rcrit(x)
\end{align*}
\end{footnotesize}
And
\begin{footnotesize}
$\B = \{\neg \G \left(\neg\rwait(x) \lor \neg\G\neg \rcrit(x)\right) \, , \, \neg \G \neg \G \rcrit(x)\}$.
\end{footnotesize}
Therefore, $\witnessvocab$ extends the original vocabulary with the following 6 unary relations:
\begin{align*}
&\relg{\neg \rscheduled(x)},
\relg{\neg \G \neg \rscheduled(x)},
\relg{\neg\rcrit(x),}\\
&\relg{\neg\rwait(x) \lor \neg\G\neg\rcrit(x)},
\relg{\rcrit(x)},
\relg{\neg \G \rcrit(x)}
\end{align*}
as well as two constants for prophecy witnesses: $c_1$ for $\neg \G \left(\neg\rwait(x) \lor \neg\G\neg \rcrit(x)\right)$, and $c_2$ for $\neg \G \neg \G \rcrit(x)$.

We now explain why there is no abstract lasso.
To do this, we show that the tableau construction, combined with the dynamic abstraction and the fair segment requirements,
result in the same reasoning that was presented informally in \Cref{sec:intro:ticket}.

First, observe that from the definition of $c_1$ and the
negation of the liveness property (both assumed by $\witnessinit$),
we have that the initial state $s_0 \models \sat{\neg \G \left(\neg\rwait(c_1) \lor \neg\G\neg \rcrit(c_1)\right)}$.
For brevity, denote $p =  \left(\neg\rwait(c_1) \lor \neg\G\neg \rcrit(c_1)\right)$, so we have
$s_0 \models \sat{\neg \G p}$, i.e., $s_0 \models \neg \relg{p}$.
Since $c_1$ is also in the footprint of the initial state, the fair segment requirement ensures that
the freeze point can only happen after encountering a state satisfying:
$\sat{(\G p) \lor \neg p} \equiv \relg{p} \lor \sat{\neg p}$.
Recall that the transition relation of the tableau ($\tabtr$),
ensures $(\relg{p}) \leftrightarrow (\sat{p} \wedge \relg{p}')$.
Therefore, on update from a state satisfying $\neg \relg{p}$ to a state satisfying $\relg{p}$
can only happen if the pre-state satisfies $\sat{\neg p}$.
Therefore, the freeze point must come after encountering a state that satisfies
$\sat{\neg p} \equiv \rwait(c_1) \land \relg{\neg\rcrit(c_1)}$.
From the freeze point onward, $\tabtr$ will ensure both $\relg{\neg\rcrit(c_1)}$ and $\neg\rcrit(c_1)$ continue to hold,
so $c_1$ will stay in $\rwait$ (since the protocol does not allow to go from $\rwait$ to anything but $\rcrit$).
So, we see that the mechanism of the tableau, combined with the temporal prophecy witness and the fair segment requirement,
ensures that the freeze point happens after $c_1$ makes a request that is never granted. This will ensure that the footprint used for the dynamic abstraction
will include all threads ahead of $c_1$ in line, i.e., those with smaller ticket numbers.

As for $c_2$, the initial state will either satisfy $\sat{\neg \G \neg \G \rcrit(c_2)}$
or it would satisfy $\sat{\neg \exists x. \neg \G \neg \G \rcrit(x)}$.
In the first case, by an argument similar to the one used above for $c_1$, the freeze point will happen
after $c_2$ enters the critical section and then stays in it. Therefore, the footprint used for the dynamic abstraction
will include all numbers smaller than $q$ of $c_2$ when it enters the critical section\footnote{When modeling natural numbers in first-order logic,
the footprint is adjusted to include all numbers lower than any constant (still being a finite set).}.
Since $c_2$ is required to be scheduled between the repeating states (again by the tableau construction and the fair segment requirement), its value for $q$ will be decreased, and this will be visible in the dynamic abstraction. Thus, in this case, an abstract lasso is not possible.

In the second case the initial state satisfies $\sat{\neg \exists x. \neg \G \neg \G \rcrit(x)}$.
By a similar argument that combines the tableau with the fair segment requirement for the repeating states,
we will obtain that between the repeating states, any thread in the footprint of the first repeating state, must both be scheduled and visit a state outside the critical section. In particular, this includes all threads that are ahead of $c_1$ in line.
This entails a change to the program counter of one of them (the one that had a ticket number equal to the service number at the first repeating state), which will be visible in the abstraction. Thus, an abstract lasso is not possible in this case either.

%\section{Completeness w.r.t. First-Order Reasoning}
\section{Closure Under First-Order Reasoning}
\label{sec:cut}

\commentout{
The reduction from temporal verification to safety verification is inherently
incomplete.  However, in this section, we show that the set of
instances for which the reduction is complete is closed under
first-order reasoning. This is unlike previous reductions. It
shows that the use of temporal prophecy results in a particular kind
of robustness, albeit not in completeness (which is impossible for a
reduction from arbitrary FO-LTL properties to safety properties).
}

The transformation from temporal verification to safety verification
developed in \Cref{sec:reduction} introduces an abstraction, and
incurs a loss of precision. That is, for some systems and properties,
liveness holds but the safety of the resulting system does not hold,
no matter what temporal prophecy is used. (This is unavoidable for a
transformation from arbitrary FO-LTL properties to safety properties~\cite{popl18}.)
However, in this section, we show that the set of instances for which
the transformation can be made precise (via temporal prophecy) is
closed under first-order reasoning. This is unlike the transformation of \cite{popl18}.
It shows that the use of temporal prophecy results in a particular kind of robustness.

We consider a proof system in which the above transformation is
performed and the resulting safety property is checked by an oracle.
That is, for a transition system $\insystem$ and a temporal property
$\varphi$ (a closed FO-LTL formula), we write
$\redproves{\insystem}{\varphi}$ if there exist finite sets of
FO-LTL formulas $\A$ and $\B$ satisfying the conditions of
\Cref{thm:soundness}, such that resulting transition system
$\witnesssystem$ is safe, i.e., does not contain an abstract lasso.
We now show that the relation $\vdash$ satisfies a powerful closure property.

\begin{theorem}[Closure under first-order reasoning]
\label{thm:closure}
Let $\insystem$ be a transition system, and $\psi,\varphi_1,\ldots,\varphi_n$
be closed FO-LTL formulas, such that
$\sat{\varphi_1 \land \ldots \land \varphi_n} \models \sat{\psi}$.
If $\redproves{\insystem}{\varphi_i}$ for all $1 \leq i \leq n$, then
$\redproves{\insystem}{\psi}$.
\end{theorem}

The condition that $\sat{\varphi_1 \land \ldots \land \varphi_n}
\models \sat{\psi}$ means that $\varphi_1 \land \ldots \land
\varphi_n$ entails $\psi$ when using only first-order reasoning, and
treating temporal operators as uninterpreted. The theorem states that
provability using the liveness-to-safety transformation is closed
under such reasoning. Two special cases of \Cref{thm:closure} given by
the following corollaries:

\begin{corollary}[Modus Ponens]
If $\insystem$ is a transition system and $\varphi$ and $\psi$ are closed FO-LTL formulas such that
$\redproves{\insystem}{\varphi}$ and $\redproves{\insystem}{\varphi \to \psi}$, then
$\redproves{\insystem}{\psi}$.
\end{corollary}

\begin{corollary}[Cut]
If $\insystem$ is a transition system and $\varphi$ and $\psi$ are closed FO-LTL formulas such that
$\redproves{\insystem}{\varphi \to \psi}$ and $\redproves{\insystem}{\neg \varphi \to \psi}$, then $\redproves{\insystem}{\psi}$.
\end{corollary}

\begin{IEEEproof}[Proof of \Cref{thm:closure}]
In the proof we use the notation $\witnesssystem(\insystem,\varphi,\A,\B)$ to denote the transition system constructed for $\insystem$ and $\varphi$ when using $\A,\B$ as temporal prophecy formulas. Likewise, we refer to the vocabulary, initial states and transition relation formulas of the transition system as  $\witnessvocab(\insystem,\varphi,\A,\B)$, $\witnessinit(\insystem,\varphi,\A,\B)$, and $\witnesstr(\insystem,\varphi,\A,\B)$, respectively.
Let $(\A_1,\B_1),\ldots,(\A_n,\B_n)$ be such that $\witnesssystem(\insystem,\varphi_i,\A_i,\B_i)$ has no abstract lasso, for every $1 \leq i\leq n$.
%the reduction for $T$ and $\varphi_i$ with $\A_i,\B_i$ does not result in an abstract lasso.
Now, let $\A = \bigcup_{i=1}^n \A_i$ and $\B = \bigcup_{i=1}^n \B_i$. We show that $\witnesssystem(\insystem,\psi,\A,\B)$ has no abstract lasso.
%the reduction for $T$ and $\psi$ with $\A,\B$ does not result in an abstract lasso,
Assume to the contrary that $s_0,\ldots,s_i,\ldots,s_j,\ldots,s_k,\ldots,s_n$ is an abstract lasso for $\witnesssystem(\insystem,\psi,\A,\B)$. %this reduction.
Since $s_0 \models \witnessinit(\insystem,\psi,\A,\B)$, we know that $s_0 \models \neg \sat{\psi}$, and since $\sat{\varphi_1 \land \ldots \land \varphi_n} \models \sat{\psi}$,
there must be some $1 \leq \ell \leq n$ s.t. $s_0 \models \neg \sat{\varphi_\ell}$. Denote $\vocabulary' = \witnessvocab(\insystem,\varphi_\ell,\A_\ell,\B_\ell)$. Now, $\proj{s_0}{\vocabulary'},\ldots,\proj{s_i}{\vocabulary'},\ldots,\proj{s_j}{\vocabulary'},\ldots,\proj{s_k}{\vocabulary'},\ldots,\proj{s_n}{\vocabulary'}$
is an abstract lasso of $\witnesssystem(\insystem,\varphi_\ell,\A_\ell,\B_\ell)$, which is a contradiction.
To see that, we first simplify the notation and denote $\proj{s_m}{\vocabulary'}$ by $\hat{s}_m$.
%\TODO{here we should really project to the vocabulary of the $v$-reduction, once we have better notations} can be projected to an abstract lasso for the
%reduction for $T,\varphi_v$ with $\A_v,\B_v$, which is a contradiction.
The footprint $f(s_0,\ldots,s_i)$ contains more elements than the footprint $f(\hat{s}_0,\ldots,\hat{s_i})$, since $\witnessvocab(\insystem,\psi,\A,\B) \supseteq  \witnessvocab(\insystem,\varphi_\ell,\A_\ell,\B_\ell)$.
Therefore, given that $\proj{{s}_j}{f({s}_0,\ldots,{s_i})} = \proj{{s}_k}{f({s}_0,\ldots,{s_i})}$, we have that
 $\proj{\hat{s}_j}{f(\hat{s}_0,\ldots,\hat{s_i})} = \proj{\hat{s}_k}{f(\hat{s}_0,\ldots,\hat{s_i})}$ as well.
%, when $D^v$ denotes the footprint in the reduction with $\A_\ell,\B_\ell$.
Moreover, the fairness constraints in  $\witnesssystem(\insystem,\varphi_\ell,\A_\ell,\B_\ell)$, determined by $\A_\ell$, are a subset of those in $\witnesssystem(\insystem,\psi,\A,\B)$), determined by $\A$, so the segments $[0,i]$ and $[j,k]$ are also fair in $\witnesssystem(\insystem,\varphi_\ell,\A_\ell,\B_\ell)$.
%To see this, note that the footprint of the reduction with $\A,\B$ contains more elements than the footprint in the reduction with $\A_\ell,\B_\ell$ (as the vocabulary contains more constants),
%so $\proj{s_j}{D^v_i} = \proj{s_k}{D^v_i}$, when $D^v$ denotes the footprint in the reduction with $\A_v,\B_v$.
%Moreover, in the reduction with $\A,\B$, there are more fairness constraints compared to the reduction with $\A_v,\B_v$, so the segments $[0,i]$ and $[j,k]$ are also fair in the latter reduction.
\end{IEEEproof}

The proof of \Cref{thm:closure} sheds more light on the power of using
temporal prophecy formulas that are not subformulas of the temporal property to prove.
In particular, the theorem does not hold if $\A$ is restricted to subformulas of the temporal proof goal.

\section{Implementation \& Evaluation}
\label{sec:eval}
%\remspace

%\subsection{Implementation in Ivy}
%\remspace

We have implemented our approach for temporal verification and integrated it into the Ivy deductive verification system~\cite{ivy}.
This allows the user to model the transition system in the Ivy language (which internally translates into a first-order transition system),
and express temporal properties directly in FO-LTL.
In our implementation, the safety property that results from the liveness-to-safety transformation is proven by a suitable inductive invariant, provided by the user.
To facilitate this process, Ivy internally constructs a suitable monitor for the safety property, i.e., the absence of abstract lasso's in $\witnesssystem$.
The user then provides an inductive invariant for $\witnesssystem$ composed with this monitor.
The monitor keeps track of the footprint and the fairness constraints,
and non-deterministically selects the freeze point and repeated states of the abstract lasso. Similar to the construction of~\cite{livenessassafety},
the monitor keeps a shadow copy of the ``saved state'', which is the first of the two repeated states.
These are maintained via designated relation symbols (in addition to $\witnessvocab$).
The user's inductive invariant must then prove that it is impossible for the monitor to detect an abstract lasso.

%\TODO{maybe mention that this monitor is explained in more detail in \cite{popl18}}

\para{Mining Temporal Prophecy from the Invariant}
As presented in previous sections, our liveness-to-safety transformation is parameterized by sets of formulas $\A$ and $\B$.
In the implementation, these sets are implicit, and are extracted automatically from the inductive invariant provided by the user. %, which may contain temporal formulas as well as prophecy witness constants.
Namely, the inductive invariant provided by the user contains temporal formulas, and also prophecy witness constants, where every temporal formula $\G \varphi$ is a shorthand (and is internally rewritten to) $\relg{\varphi}$.
%It is internally rewritten to use $\relg{\varphi}$ instead of $\G \varphi$, for every temporal formula.
%
The set $\A$ to be used in the construction is defined by all the temporal subformulas that appear in the %user provided
inductive invariant (and all their subformulas), and the set $\B$ is defined according to the prophecy witness constants that are used in the inductive invariant.
%The inductive invariant itself is internally rewritten to use $\relg{\varphi}$ instead of $\G \varphi$, for every temporal formula used in the invariant.
%\oded{Maybe remove this or move somewhere else: As an inductive invariant describes a property of the reachable states (and not of traces), the user should think of the temporal formulas in the invariant as anticipating the future behavior of the outgoing traces.}

In particular, the user's invariant may refer to the satisfaction of each fairness constraint $\sat{\G \varphi \lor \neg \varphi}$
 for $\G \varphi \in \A$, both before the freeze point and between the repeated states, via a convenient syntax provided by Ivy.

\commentout{
In addition to symbols in $\witnessvocab$, the monitor contains relations that keep track of the footprint,
the saved state, \sharon{the rest only explains the fairness. So can we avoid mentioning the footprint and saved state here?}
and the satisfaction of the fairness constraints $\sat{\G \varphi \lor \neg \varphi}$ for $\G \varphi \in \A$, both before the freeze point and
between the repeated states. The inductive invariant provided by the user must refer to these relations as well.
This is enabled by a convenient syntax, that allows the user to refer to each fairness constraint by the formula $\varphi$, associated with it.
}

\para{Interacting with Ivy}
If the user provides an inductive invariant that is not inductive, Ivy presents a graphical counterexample to induction.
This guides the user to adjust the inductive invariant, which may also lead to new formulas being added to $\A$ or $\B$,
if the user adds new temporal formulas or prophecy witnesses to the inductive invariant.
In this process, the user's mental image is of a liveness-to-safety transformation where $\A$ and $\B$ include all (countably many) FO-LTL formulas over the system's vocabulary,
so the user is free to use any temporal formula, or prophecy witness for any formula.
However, since the user's inductive invariant is a finite formula, the liveness-to-safety transformation needs only to be applied to finite $\A$ and $\B$,
and the infinite $\A$ and $\B$ are just a mental model.

\TODO{should we say something about decidability?}

\TODO{syntax for invariants that refer to the reduced system}

\TODO{maybe mention ``normalization'' of formulas}

\TODO{Can we give the simplest possible example of an invariant? Otherwise this seems like magic.}

\TODO{Add this somehow:
Q: In the evaluation, could you explain why temporal prophecy appears to need more conjectures than the POPL'18 work for the ABP and TLB shootdown examples?
A: In POPL18, temporal monitors were constructed manually, and contained some shortcuts. In contrast, the tableau construction, which is made automatic in this work (and implemented in Ivy), results in systems which require slightly more verbose invariants. E.g., for ticket, FO( forall x.[]<> scheduled(x) ) must be included in the invariant to assert that it keeps holding. In contrast, in POPL18, scheduled(x) was taken as an apriori fairness constraint (a manual optimization of the tableau construction). In ABP, some conjectures related to fairness constraints repeat 4 times (for combinations of sender,receiver bits).}

%\subsection{Experiments}
%\remspace

\commentout{
|A|, |B|, #LOC of model, # conjectures, # conjectures that use temporal operators
}

\begin{figure}
  \begin{footnotesize}
  \arraycolsep=3pt
  \[
  \begin{array}{||l|c|c|c|c|c|c||}
    \hhline{|t:=======:t|}
    \textbf{Protocol} &
    \textbf{\# A} &
    \textbf{\# B} &
    \textbf{\# LOC} &
    \textbf{\# C} &
    \textbf{FO-LTL} &
    \mathbf{t} \text{ [sec]}
    \\
    \hhline{||-------||}
    \mbox{Ticket w/ Task Queues}
    &
    1
    &
    2
    &
    90
    &
    60
    &
    22\%
    &
    9.4
    \\
    \hhline{||-------||}
    \mbox{Alternating Bit Protocol}
    &
    4
    &
    1
    &
    143
    &
    70
    &
    40\%
    &
    32
    \\
    \hhline{||-------||}
    \mbox{TLB Shootdown}
    &
    6
    &
    3
    &
    468
    &
    102
    &
    49\%
    &
    283
    \\
    \hhline{|b:=======:b|}
  \end{array}
  \]
  \end{footnotesize}
  \caption{ \label{fig:eval}%
    Protocols for which we verified liveness.
    For each protocol, $\textbf{\# A}$ reports the number of temporal prophecy formulas used.
    $\textbf{\# B}$ reports the number of prophecy witnesses used.
    $\textbf{\# LOC}$ reports the number of lines of code for the system model (without proof) in Ivy's modeling language.
    $\textbf{\# C}$ reports the number of conjectures used in the inductive invariant (a typical conjecture is one or few lines).
    $\textbf{FO-LTL }$ reports the fraction of the conjectures that use temporal formulas.
    Finally, $\textbf{t}$ reports the run time (in seconds) for checking the  verification conditions using Ivy and Z3.
    The experiments were performed on a laptop running 64-bit Linux,
    with a Core-i7 1.8 GHz CPU, using Z3 version 4.6.0.
    \TODO{mention commit (after we merge to master}
    \TODO{bring this back: Z3 uses heuristics which employ randomness. Therefore, each experiment was repeated 10 times using random seeds, and we report the mean verification time.}
  }
\end{figure}

We have used our implementation to prove liveness for several challenging examples, summarized in \Cref{fig:eval}.
We focused on examples that were beyond reach for the liveness-to-safety transformation of~\cite{popl18}.
In~\cite{popl18}, such examples were handled using a nesting structure.
Our experience shows that with temporal prophecy, the invariants are simpler than with a nesting structure (for additional comparison with nesting structure see \Cref{related}).
For all examples we considered, the verification conditions are in a decidable fragment of first-order logic which is supported by Z3
(the stratified extension of EPR \cite{ivy,ge2009complete}).
Interestingly, for the TLB shootdown example, the proof presented in~\cite{popl18} (using a nesting structure) required non-stratified quantifier alternation,
which is eliminated by the use of temporal prophecy witnesses.
Due to the decidability of verification conditions, Z3 behaves predictably, and whenever the invariant is not inductive it
produces a finite counterexample to induction, which Ivy presents graphically.
Our experience shows that the graphical counterexamples provide valuable guidance towards finding an inductive invariant,
and also for coming up with temporal prophecy formulas as needed.
Below we provide more detail on each example.

%\TODO{what else should we say here?}
%\TODO{Describe each of the three examples and say what the key predicates were in the invariants that made the proofs work. Even better would be to say how the counterexamples to induction helpd to determine what addition predicates were needed in the invariants.}
%\TODO{Say what the columns in the table mean.}

%\subsubsection{Ticket}
\para{Ticket}
The ticket example has been discussed in \Cref{sec:intro}, and
\Cref{sec:ticket} contains more details about its proof with temporal
prophecy, using a single temporal prophecy formula and two prophecy witness constants.
To give a flavor of what the proof looks like in Ivy, we present a couple of the
conjectures that make up the inductive invariant for the resulting system, in Ivy's syntax.
In Ivy, the prefix \texttt{l2s} indicates symbols that are introduced by the liveness-to-safety transformation.
\commentout{
Some of the conjectures mention only properties of reachable states of
the original system that are important for liveness. An example of this is:
\begin{footnotesize}
\begin{verbatim}
forall K. n > K & s <= K -> exists T. m(T) = K & ~idle(T)
\end{verbatim}
\end{footnotesize}
This conjecture states that for any ticket number between $s$ and $n$,
there is a thread holding that ticket that is also not in the $\ridle$
state (see \Cref{fig:ticket}).
}
Some conjectures are needed to state that the footprint used in the dynamic abstraction contains enough elements.
An example of such a conjecture is:

\begin{footnotesize}
\begin{verbatim}
l2s_frozen & (globally critical(c2)) ->
    forall N.  N <= q(c2) -> l2s_a(N)
\end{verbatim}
\end{footnotesize}
This conjecture states that after the freeze point (indicated by the special symbol \texttt{l2s\_frozen}),
if the prophecy witness \texttt{c2} (which is the prophecy witness defined for $\F \G \rcrit(x)$) is globally in the critical section,
then the finite domain of the frozen abstraction (stored in the unary relation \texttt{l2s\_a}) contains all numbers up the \texttt{c2}'s value for \texttt{q}.
Other conjectures are needed to show that the current state is different from the saved state.
One example is:
\begin{footnotesize}
\begin{verbatim}
l2s_saved & (globally critical(c2)) &
    ~($l2s_w X. scheduled(X))(c2) ->
        q(c2) ~= ($l2s_s X. q(X))(c2)
\end{verbatim}
\end{footnotesize}
The special operator \texttt{\$l2s\_w} lets the user query whether a fairness constraint has been encountered,
  %is exposed to the user with ,
and \texttt{\$l2s\_s}  exposes to the user the saved state
%and the saved state is exposed via \texttt{\$l2s\_s}
(both syntactically $\lambda$-like binders).
This conjecture states that after we saved a shadow state (indicated by \texttt{l2s\_save}),
if the prophecy witness \texttt{c2} is globally in the critical section,
and if we have encountered the fairness constraints associated with
$\rscheduled(x) \lor \G \neg \rscheduled(x)$ instantiated for \texttt{c2} (which can only happen after \texttt{c2} has been scheduled),
then the current value \texttt{c2} has for \texttt{q} is different from the same value in the shadow state.
\para{Alternating Bit Protocol}
The alternating bit protocol is a classic communication algorithm for
transition of messages using lossy first-in-first-out (FIFO) channels.
The protocol uses two channels: a data channel from the sender to the
receiver, and an acknowledgment channel from the receiver to the
sender.  The sender and the receiver each have a state bit, and
messages include a bit that functions as a ``sequence number''.  We
assume that the sender has an (infinite) array of values to send,
which is filled by some independent process.  The liveness property we
wish to prove is that every value entered into the sender array is
eventually received by the receiver.

The protocol is live under fair scheduling assumptions, as well as
standard fairness constraints for the channels: if messages are
infinitely often sent, then messages are infinitely often received.
This makes the structure of the temporal property more
involved. Formally, the liveness property we prove is:
\begin{small}
\begin{align*}
  &
  (\G \F \textit{sender\_scheduled}) \land
  (\G \F \textit{receiver\_scheduled}) \, \land \\
  &
  \left((\G \F \textit{data\_sent}) \to (\G \F \textit{data\_received})\right) \land \\
  &
  \left((\G \F \textit{ack\_sent}) \to (\G \F \textit{ack\_received})\right)
  \, \to \\
  &
  \forall x. \G (\textit{sender\_array}(x) \neq \bot \to \F \textit{receiver\_array}(x) \neq \bot))
\end{align*}
\end{small}
This property cannot be proven without temporal prophecy.
However, it can be proven using 4 temporal prophecy formulas:
\begin{small}
$%\begin{equation*}
\left\{ \F\G\, ( \textit{sender\_bit}=s \land \textit{receiver\_bit}=r ) \mid s,r \in \{0,1\} \right\}
$\end{small}.
%\end{equation*}
Intuitively, these formulas make a distinction between
traces in which the sender and receiver bits eventually become fixed,
and traces in which they change infinitely often.

%\subsubsection{TLB Shootdown}
%\label{sec:tlb}
\para{TLB Shootdown}
%\oded{I shortened. Jochen, please check if it's still correct}
%copied from Popl18, oded: modified
The TLB shootdown algorithm~\cite{Black:1989:TLB:70082.68193}
is used (e.g. in the Mach operating system)
to maintain consistency of Translation Look-aside Buffers (TLB) across processors.
When some processor (dubbed the initiator) changes the page table, it interrupts all other
processors currently using the page table (dubbed the responders) and
waits for them to receive the interrupt before making changes.
\commentout{
The algorithm itself runs in four phases: (1) the initiator interrupts
all processors using a page table, (2) the interrupted processors set
a flag that they are deactivated, (3) when every processor set the
flag, the initiator changes the page table and finishes, (4) the
responders can then continue and flush their TLB.  The algorithm is
further complicated by the fact that a processor can
non-deterministically choose to act as initiator in which case it
does not respond to interrupts.
}
The liveness property we prove is that no processor can become stuck either as
an initiator or as a responder (formally, it will respond or initiate infinitely often). This liveness depends on fair
scheduling assumptions, as well as strong fairness assumptions for the
page table locks used by the protocol.
We use one witness for the process that does not satisfy the liveness
property.  Another witness is used for a pagemap that is never
unlocked, if this exists.  A third witness is used for a process that
possibly gets stuck while holding the lock blocking the first process.
We use six prophecy formulas to case split on when some process may
get stuck. Two of them are used for the two loops in the initiator to
distinguish the cases whether the process that hogs the lock gets
stuck there.  They are of the form $\F \G pc(c_2)\in\{i_3,\dots,i_8\}$.
Two are used for the two lock instructions to indicate
that the first process gets stuck: $\F\G pc(c_1) = i_2$.
And two are used for the second and third witness to indicate whether
such a witness exists, e.g., $\F\G plock(c_3)$.
% We show for each witness that it leads to a terminating run.
%, i.e., it is not possible to satisfy the corresponding prophecy formula.
% oded: the above was unclear to me so I commented it out
Compared to the proof of~\cite{popl18}, our proof is simpler due to the temporal prophecy,
and avoids non-stratified quantifier alternation, resulting in decidable verification conditions.

%\remspace
\section{Related Work}
\label{related}
%\remspace

\oded{Added the first paragraph. Please check}
Prophecy variables were first introduced
in~\cite{DBLP:journals/tcs/AbadiL91}, in the context of refinement
mappings. There, prophecy variables are required to range over a
finite domain to ensure soundness. Our notion of prophecy via
first-order temporal formulas and witness constants does not meet this
criterion, but is still sound as assured by \Cref{thm:soundness}.
In~\cite{Kesten:2002:NIA:646737.701938}, LTL formulas are used to
define prophecy variables in a way that is similar to ours, but only
to show refinement between finite-state processes. We use temporal prophecy defined by FO-LTL
formulas in the context of infinite-state systems. Furthermore, we consider a
liveness-to-safety transformation (rather than refinement mappings), which
can be seen as a proof system for FO-LTL.
%We use temporal prophecy defined by FO-LTL
%formulas (that model infinite-state systems). We also consider a
%liveness-to-safety reduction (rather than refinement mappings), which
%can be seen as a proof system for FO-LTL.
%\TODO{For the final version, read \cite{Kesten:2002:NIA:646737.701938} and make sure}

The liveness-to-safety transformation based on dynamic abstraction, but \emph{without temporal prophecy}, was introduced in~\cite{popl18}.
%However, they did not use temporal prophecy.
%They used a \emph{nesting structure} to increase the power of the reduction.
There, a \emph{nesting structure} was used to increase the power of the transformation.
A nesting structure is defined by the user (via first-order formulas),
and has the effect of splitting the transition system into levels (analogous to nested loops) and proving each level separately.
Temporal prophecy as we introduce here is more general, and in particular, any proof that is possible with a nesting structure,
is also possible with temporal prophecy (by adding a temporal prophecy formula $\F \G \delta$ for every nesting level, defined by $\delta$).
Moreover, the nesting structure does not admit cut elimination or closure under first-order reasoning,
and is therefore less robust.

One effect of prophecy is to split cases in the proof on some aspect
of the future.  This very general idea occurs in various approaches to
liveness, particularly in the large body of work on lexicographic or
disjunctive rankings for
termination~\cite{conf/sas/Urban13,conf/sas/UrbanM14,conf/tacas/UrbanGK16,sefm/BabicHRC07,conf/tacas/CookSZ13,popl/CousotC12,cav/GantyG13,rta/GieslTSF04,conf/pldi/GrebenshchikovLPR12,conf/esop/UrbanM14,conf/cav/BrockschmidtCF13,pldi/CookPR06,cook-terminator,sas/HarrisLNR10,cav/KroeningSTW10,cav/LeeWY12,lics/PodelskiR04,popl/PodelskiR05}.
In the work of \cite{pldi/GulwaniJK09}, the partitioning of the
space of potentially infinite executions is based on the \emph{a
priori} decomposition of regular expressions for iterated loop
segments. Often the partitioning here amounts to a split according
to a fairness condition (``command $a$ is taken infinitely often or
it is not''). The partitioning is constructed dynamically (and
represented explicitly through a union of Buchi automata) in
\cite{conf/cav/HeizmannHP14} (for termination),
in~\cite{DBLP:conf/cav/DietschHLP15} (for liveness), and
in~\cite{DBLP:conf/lics/FarzanKP16} (for liveness of parameterized
systems).  None of these works uses a temporal tableau construction
to partition the space of futures, however.

% Our work can be viewed as one possible way to realize the idea of splitting the
% space of potentially infinite executions into independent cases for
% proofs of termination and liveness properties.  In this rather wide
% sense, it is related to a huge body of research on lexicographic or
% disjunctive proofs for termination and
% liveness; see,
% % hoping that the references get ordered automatically!!!
% e.g.,~\cite{conf/sas/Urban13,conf/sas/UrbanM14,conf/tacas/UrbanGK16, sefm/BabicHRC07,conf/tacas/CookSZ13,popl/CousotC12,cav/GantyG13,rta/GieslTSF04,conf/pldi/GrebenshchikovLPR12,conf/esop/UrbanM14,conf/cav/BrockschmidtCF13,pldi/CookPR06,cacm/CookPR11,sas/HarrisLNR10,cav/KroeningSTW10,cav/LeeWY12,lics/PodelskiR04,popl/PodelskiR0}.  Specifically, in
% the work of \cite{pldi/GulwaniJK09}, the partitioning of the space of
% potentially infinite executions is based on the \emph{a priori} decomposition of
% regular expressions for iterated loop segments.  Often the
% partitioning here amounts to a split according to a fairness condition
% (``command $a$ is taken infinitely often or it is not'').   The partitioning is constructed
% dynamically (and represented explicitely through a union of Buchi
% automata) in
% \cite{conf/cav/HeizmannHP14} (for termination),
% in~\cite{DBLP:conf/cav/DietschHLP15} (for liveness), and
%  in~\cite{DBLP:conf/lics/FarzanKP16} (for liveness of parametrized systems).

Here, we use prophecy to, in effect, partially determinize a system by making non-deterministic choices earlier in an execution. This same effect was used for a different purpose in refining an abstraction from LTL to ACTL~\cite{DBLP:conf/popl/CookK11} and checking CTL* properties~\cite{DBLP:conf/cav/CookKP15}. The prophecy in this case relates only to the next transition and is not expressed temporally. The method of ``temporal case splitting'' in~\cite{DBLP:journals/scp/McMillan00} can also be seen as a way to introduce prophecy variables to increase the precision of an abstraction, though in that case the transformation was to finite-state liveness, not infinite-state safety. Moreover, it only introduces temporal witnesses.

\commentout{
\oded{added this, but not sure if this should be here or somewhere else in this section}
Recently, an approach based on lasso detection has been developed for
runtime detection of liveness
bugs~\cite{DBLP:conf/fmcad/MudduluruDDLQ17}.
}

We have considered only proof methods that transform
liveness to safety (which includes the classical ranking approach for
while loops). There are approaches, however, which do \emph{not}
transform liveness to safety.  For example, the approaches in
\cite{DBLP:conf/concur/AbdullaJRS06,popl/CousotC12,DBLP:journals/cl/UrbanM17}
are essentially forms of widening in a CTL-style backwards fixpoint
iteration. It is not clear to what extent temporal prophecy might be
useful in increasing the precision of such abstractions, but it may be
an interesting topic for future research.

% In
% \Cref{sec:eval} we will discuss more technical details regarding the
% work in~\cite{DBLP:conf/concur/AbdullaJRS06}.  The approaches have not
% yet been extended to deal with general liveness and fairness and to
% deal with the data structures typically found in distributed
% protocols.  In the termination analysis for heap-manipulating programs
% presented in~\cite{DBLP:conf/cav/ManevichDR16}, the use of ranking
% functions is avoided by a novel form of reasoning over sets of nodes,
% sets which are \emph{a priori} known to be finite.  In this aspect,
% the work is related to ours (perhaps more spiritually than
% technically).

%\vspace{-0.3cm}
\section{Conclusion}
%\vspace{-0.3cm}

We have seen that the addition of prophecy variables in the form of
temporal formulas can increase the precision of liveness-to-safety
tranformations for infinite-state systems. The prophecy variables are
derived from additional temporal formulas that in our implementation
were mined from the invariants a user provides to prove the safety
property. This approach is effective for proving challenging examples.
By increasing the precision of the dynamic abstraction, it avoided the
need to decompose the proof into nested termination arguments,
reducing the human effort of proof construction. Though completeness
is not possible, we saw that the additional expressiveness of temporal
prophecy provides a cut elimination property.  While we considered
temporal prophecy using a particular liveness-to-safety construction
(based on dynamic abstraction), it seems reasonable to expect that the
tableau-based approach would apply to other constructions and
abstractions, including constructions based on rankings and
well-founded relations.  Because our approach relies on an inductive
invariant supplied by the user, it requires the user to understand the
liveness-to-safety transformation and it requires both cleverness and
a deep understanding of the protocol. For this reason, a possible
avenue for future research would be to explore invariant synthesis
techniques, and in particular ones that account for refinement due to
temporal prophecy.

\TODO{something more about prophecy witnesses breaking quantifier alternation cycles?}

\label{sec:conclusion}

%\begin{acks}
\remspace
\subsubsection*{Acknowledgements}
We thank
%%
%% Thomas Ball,
%Amir Ben-Amram,
%% Nikolaj Bj{\o}rner,
%% Tej Chajed,
%% Constantin Enea,
%% Yotam M. Y. Feldman,
%Neil Immerman,
%% Daniel Jackson,
%% Ranjit Jhala,
%% K. Rustan M. Leino,
%% Giuliano Losa,
%Ken McMillan,
%% Yuri Meshman,
%Alexander Nutz,
%% Bryan Parno,
%% Shaz Qadeer,
%Alexander Rabinovich,
%% Zachary Tatlock,
%% James R. Wilcox,
%%
%our shepherd Grigore Rosu,
the anonymous referees
%and the anonymous artifact evaluation referees
for insightful comments which improved this paper.
%\TODO{anyone else we should thank?}
%
Padon was supported by Google under a PhD fellowship.
Padon and Sagiv were supported by the European Research
Council under the European Union's Seventh Framework Program
(FP7/2007--2013) / ERC grant agreement no. [321174-VSSC].
This publication is part of a project that has received funding from the European Research Council (ERC) under the European Union's Horizon 2020 research and innovation programme (grant agreement No [759102-SVIS]).
The research was partially supported by
Len Blavatnik and the Blavatnik Family foundation, and by the Blavatnik Interdisciplinary Cyber Research Center, Tel Aviv University.
This material is based upon work supported by
%\TODO{is the NSF relevant?} the National Science Foundation under Grant No. 1655166
%and by
the United States-Israel Binational Science Foundation (BSF) grants No. 2016260 and 2012259.
%
%\TODO{More grants we should mention?}
%\end{acks}

%% Bibliography

% trigger a \newpage just before the given reference
% number - used to balance the columns on the last page
% adjust value as needed - may need to be readjusted if
% the document is modified later
%\IEEEtriggeratref{8}
% The "triggered" command can be changed if desired:
%\IEEEtriggercmd{\enlargethispage{-5in}}

% references section

% can use a bibliography generated by BibTeX as a .bbl file
% BibTeX documentation can be easily obtained at:
% http://mirror.ctan.org/biblio/bibtex/contrib/doc/
% The IEEEtran BibTeX style support page is at:
% http://www.michaelshell.org/tex/ieeetran/bibtex/
% \bibliographystyle{IEEEtran}
\bibliographystyle{IEEEtranS} % A version of IEEEtran.bst that sorts the entries.
% argument is your BibTeX string definitions and bibliography database(s)
\bibliography{IEEEabrv,refs}

%\newpage
%\appendix
%\input{ticket-fail}
% \input{ticket-full}

\end{document}